\documentclass[10pt,journal,compsoc]{IEEEtran}

\ifCLASSOPTIONcompsoc
  \usepackage[nocompress]{cite}
\else
  \usepackage{cite}
\fi

\ifCLASSINFOpdf
  \usepackage[pdftex]{graphicx}
\else
  \usepackage[dvips]{graphicx}
\fi

\usepackage{times}
\usepackage{subfigure}
\usepackage{booktabs} 
\usepackage{amsmath}
\usepackage{float}
\usepackage{amsthm}
\usepackage{balance}
\usepackage{verbatim}
\usepackage[keeplastbox]{flushend}
\usepackage{placeins,xspace}

\usepackage{algorithm} 
\usepackage{algpseudocode}

\usepackage{hyperref}
\usepackage{amsfonts}
\usepackage{amssymb}
\usepackage{color}
\usepackage{multirow}

\newcommand{\E}{\mathbb{E}}

\newcommand{\tb}{\mathrm{\psi}}
\usepackage{bbm}
\usepackage{float}

\newcommand{\hl}[1]{{ #1}}

\makeatletter
\newcommand{\plusline}{%
  \let\old@ALC@lno=\ALC@lno%
  \renewcommand{\ALC@lno} {+%
    \global\let\ALC@lno=\old@ALC@lno}%
}
\makeatother

\newcommand{\binsketch}{\mathrm{BinSketch}\xspace}

\newcommand{\deepwalk}{\texttt{deepwalk}\xspace}
\newcommand{\ntv}{\texttt{node2vec}\xspace}
\newcommand{\LINE}{\texttt{LINE}\xspace}
\newcommand{\quint}{{\tt QUINT}\xspace}

\newcommand{\DNS}[1][]{{\tt DNS}}
\newcommand{\OOM}{{\tt OOM}}
\newcommand{\NA}[1][]{$-$}

\usepackage{array}
\newcolumntype{H}{>{\setbox0=\hbox\bgroup}c<{\egroup}@{}}

  \newtheorem{theorem}{Theorem}
  \newtheorem{lemma}[theorem]{Lemma}

\newtheorem{obs}[theorem]{Observation}

\newtheorem{claim}[theorem]{Claim}
\newenvironment{subproof}[1][\proofname]{%
  \begin{proof}[#1]%
}{%
  \end{proof}%
}

\begin{document}
\title{\textbf{\texttt{QUINT}}: Node embedding using network hashing}

\author{Debajyoti~Bera,
    Rameshwar~Pratap,
    Bhisham Dev Verma,
        Biswadeep~Sen,
	and~Tanmoy~Chakraborty
\IEEEcompsocitemizethanks{
\IEEEcompsocthanksitem D.~Bera and T.~Chakraborty are with IIIT-Delhi, New Delhi, India.
\IEEEcompsocthanksitem R.~Pratap and B.D.~Verma are with Indian Institute of Technology, Mandi, Himachal Pradesh, India.
\IEEEcompsocthanksitem B.~Sen is currently a research associate at department of Computer Science, National University of Singapore, Singapore. This  work was done when he was affiliated with Chennai Mathematical Institute, (CMI). 

\IEEEcompsocthanksitem D. Bera and R. Pratap are equal contributors.
}
\thanks{The current paper is an extension of our recent study \cite{binsketch} on a binary sketching.}}

\IEEEtitleabstractindextext{%
\begin{abstract}
Representation learning using network embedding has received tremendous attention due to its efficacy to solve downstream tasks. Popular embedding methods (such as \deepwalk, \ntv, \LINE) are based on a neural architecture, thus unable to scale on large networks both in terms of time and space usage. 
Recently, we proposed $\binsketch$, a sketching technique for compressing binary vectors to binary vectors. 
%
In this paper, we show how to extend $\binsketch$ and use it for network hashing. Our proposal named \quint\ is built upon $\binsketch$, and it embeds nodes of a sparse network onto a low-dimensional space using simple bit-wise operations.
{\tt QUINT} is the first of its kind that provides  tremendous gain in terms of speed and space usage without compromising much on the accuracy of the  downstream tasks.  
Extensive experiments are conducted to compare {\tt QUINT} with seven state-of-the-art network embedding methods for two end tasks -- link prediction and node classification. We observe huge performance gain for {\tt QUINT} in terms of speedup (up to $7000\times$  
and space saving (up to {$80\times$})
due to its bit-wise nature to obtain node embedding. Moreover, {\tt QUINT} is a consistent top-performer for both the tasks among the baselines across all the datasets.
Our empirical observations are backed by rigorous theoretical analysis to justify the effectiveness of {\tt QUINT}. In particular, we prove that {\tt QUINT} retains enough structural information which can be used further to approximate many topological properties of networks with high confidence.
\end{abstract}

}
\maketitle
\IEEEdisplaynontitleabstractindextext

%
\IEEEpeerreviewmaketitle

 \section{Introduction}

Machine learning tasks that involve networks, such as node classification, clustering, link prediction, find it challenging to use large networks due to the difficulty in succinctly representing their structures. For example, representing the neighborhood information of a node in an $n$-node network would require an $n$-dimensional binary vector or a $k$-dimensional real vector in which $k$ indicates the number of neighbors of any node, and each entry takes $\log n$ bits. This motivated the problem of {\em network embedding}, also known as {\em network representation learning}, in which each node is assigned a low-dimensional vector such that it maintains the structural or geometric relationship among the nodes.

A number of challenges have been uncovered in the last few decades that have led to a series of studies to propose  efficient and effective node embedding methods ~\cite{HamiltonYL17}. Most embedding techniques ``learn'' an optimal encoding for specific machine learning tasks. {This involves employing an optimization algorithm (say, gradient descent) which usually lacks worst-case guarantees and is also difficult to scale for massive networks}. Recently, a few GPU-based approaches were proposed to improve scalability~\cite{gpu_embedding}. However, we are interested in CPU-only approaches. Furthermore, the embeddings obtained thus are optimized for specific tasks and may not be effective for others. The final challenge is the space complexity of the embeddings. It is desirable to get a succinct embedding of nodes because of two reasons: (i) It requires less storage; e.g., employing 128 dimensional real-valued vectors costs 1KB of storage per node (on a 64-bit machine) which shoots to 1GB for a million node network. (ii) A succinct embedding leads to faster training and inferencing of machine learning models.




Preserving the semantics of latent representation is another challenge since many proposed techniques are evaluated using statistical methods on standard datasets which may fail to bring out the rich structural invariants that these techniques might be capturing. 
A few approaches such as {\tt deepwalk} that captures short random walks~\cite{deepwalk} and {\tt GraRep} that captures powers of an adjacency matrix~\cite{GraRep2015}, explicitly consider certain structural aspects.
{However, to the best of our knowledge, how their embeddings fare on other structural properties of a graph is yet to be established.}

Of late, randomization has been proved to be highly effective in algorithm design; { yet, barring a few approaches based on random walks~\cite{deepwalk,node2vec}, random projections~\cite{Zhang2018BillionScaleNE}, and random hashing~\cite{nethash}, randomized approaches are not extensively studied in the domain of graph embedding.} {Walking along the relatively less popular path of  non-learning based and randomized approaches}, we present {\tt QUINT}, a {\bf QU}ick th{\bf IN} and bi{\bf T}ty mapping of nodes in a sparse network to low-dimensional bit-vectors. The prime takeaway from this work is { the affirmation of bit-wise hashing as a sound alternative to learning based methods. It has theoretical bounds on the worst-case situations, delivers comparable (and sometimes better) accuracy with much less space and requires significantly less time compared to the state-of-the-art.} This work also suggests that sparse data can be analysed better with sparsity-aware methods; {despite many networks being sparse in nature, we are aware of only a handful of approaches specifically for them~\cite{netsmf,strap,SSNE}}.\\

\noindent{\large \textbf{Major contributions:}}
Our main contribution is \texttt{QUINT} that takes a large $n$-node network as an input and outputs a {\em $d$-bit binary embedding} corresponding to each node. \quint uses a modified version of $\binsketch$ \cite{binsketch}, a sketching algorithm that we recently proposed for binary data.
Here, $d$ denotes the dimension of the embedding, and we give a formula to compute it in terms of the maximum degree of any node of a network (which we will simply denote by `degree' when the context is clear). We make two important modifications to $\binsketch$, one of them being a larger dimension than what is stipulated by $\binsketch$; however, it is common for real-world networks to have a low degree compared to the number of nodes, and the best part is that \hl{a much smaller $d$ suffices in practice.}
%
%
%

{Our approach addresses many of the concerns raised above and emerges as a strong contender for node embedding in a resource-constrained scenario --- it requires shorter embedding time and generates small-sized embeddings.} We emphasise that \quint is designed to handle networks {\em without attributes and weights}.

%
%

{\bf (i) Speedup in embedding:}
{\tt QUINT} takes an adjacency matrix or an edge list of a network as an input and outputs binary embedding of nodes using just one pass over the network while doing bit manipulations. Due to its simplicity, it computes the embeddings in almost real time. For instance, on the \texttt{BlogCatalog} network~\cite{BlogCatalog} with $10,312$ nodes and $333,983$ edges, \texttt{QUINT} finishes within $1.6$ secs (for $d=1000$), about $5~\mu sec$ per edge; in contrast, {\tt node2vec}, \deepwalk{}, \LINE  require 23 mins, 19 mins and 32 mins, respectively, for a single epoch.  {Similarly, on \texttt{Flickr} \cite{dataset} network  with $80,513$ nodes and $5,899,882$ edge,  \texttt{QUINT} finished within $30.63$ secs (for $d=1000$), by taking about $5.2~\mu sec$ per edge; on the other hand, {\tt node2vec} and   \LINE did not finished in  $10$ hours;  \deepwalk{}, {\tt VERSE} and {\tt NodeSketch}   require $2.1$ hr, and  $4.8$ hr and $18$ mins, respectively. Also on \texttt{Youtube} network \cite{dataset} with  $1,138,499$ nodes and $2,990,443$ edges,  \texttt{QUINT} finished within $35.06$ secs (for $d=1000$), about $11.7~\mu sec$ per edge; whereas, {\tt node2vec} , \deepwalk{}, \LINE, {\tt VERSE} did not finished in  $10$ hours and {\tt NodeSketch}   require $15$ mins.
}

{\bf (ii) Less space overhead:}
Popular  embedding methods output real-valued embedding, generally $128$ or $256$ dimensional real-valued vectors, whereas \texttt{QUINT} generates binary embeddings consuming $d$ bits for a $d$-dimensional embedding of a node. In the link prediction task { ({\it aka.} link discovery)} on \texttt{BlogCatalog}, we find that {\tt node2vec}, {\tt NodeSketch}, {\tt LINE} and {\tt deepwalk} generate $128$-dimensional real-valued vectors that consume $128\times 64=8{,}192$ bits, and these embeddings generate a maximum AUC score of $66$. In contrast, {\tt QUINT} with $d=1000$ uses only a tenth of that space (1000 bits) to achieve $84.4$ AUC-ROC. {Similarly, on Youtube dataset {\tt node2vec} and {\tt NodeSketch} for $128$-dimension embedding achieve a maximum AUC score of $65.1$, whereas  {\tt QUINT} with $d=100$ achieve $90$ AUC-ROC value, while using  only $1/80$ of space in comparison.}

{\bf (iii) State-of-the-art accuracy:}
We compare {\tt QUINT} with seven recently proposed state-of-the-art embedding methods -- {\tt node2vec}~\cite{node2vec}, {\tt deepwalk}~\cite{deepwalk}, {\tt LINE}~\cite{line}, {\tt Verse}~\cite{verse}, {\tt NetMF}~\cite{netmf}, {\tt NodeSketch}~\cite{nodesketch}, and {\tt SGH}~\cite{sgh}, for link prediction and node classification, and on altogether ten real-world publicly available datasets and {  several LFR generated synthetic graphs}. For our experiments we perform the end tasks considering the embeddings obtained from the competing methods. 
 {\tt QUINT} remains one of the top performers in terms of accuracy across all the networks and end tasks. We attribute this to $\binsketch$ that has displayed the ability to generate succinct representations which simultaneously ``preserve'' multiple similarities~\cite{binsketch}.

{\bf (iv) Supporting theoretical analysis:}
We also present a rigorous theoretical analysis of the appealing properties that give \quint an edge over the existing techniques.
Embedding obtained from {\tt QUINT} is essentially a compressed form of the neighborhood information of a node. We show that the embedding, even though compressed, retains information about the network structure such as the degrees, the number of common neighbors, the number of even length paths, etc. More generally, we discuss how any even power of an adjacency matrix can be approximately computed from the embedding with a large confidence and accuracy.
{  Not surprisingly, our embedding allows us to perform link prediction and node classification as good as the others, and sometimes better. We hope that similar performance would be observed for other machine learning tasks (e.g., clustering) that rely only on  the structural properties of a network, i.e., without using any metadata.}

{\bf (v) Sparsity-aware design:}
Many real-world networks are sparse in nature. {  However, to the best of our knowledge, most of the embedding approaches do not tend to include optimisations specific to sparse data or provide theoretical guarantees that depend on its sparsity, except a few, e.g., NetSMF~\cite{netsmf}, STRAP~\cite{strap}, SSNE~\cite{SSNE}}. \quint is designed keeping sparse networks in mind. Both the empirical and analytical results explain how \quint leverages the notion of sparsity of a network, which we quantify by the largest degree among all nodes.
 
{\bf (vi) {\tt QUINT} {\it vs.} uncompressed representation:}
Since \quint emdeddings are compressed forms of rows of an adjacency matrix, one may wonder about the payoffs when the uncompressed adjancency lists or the adjacency rows (rows of an adjacency matrix) are instead used as ``embeddings''. Adjacency lists of sparse networks are no doubt compact, but are not immediately suitable due to unequal number of neighbors, complete dependence of the embedding on the ordering of nodes, etc. Adjacency rows avoid these problems but are verbose; nevertheless we compare it with \quint. Quite surprisingly, we find that the classification and prediction performances of {\tt QUINT}  almost match the accuracy using uncompressed adjacency matrix. In fact, on {\tt Enron}, {\tt QUINT} achieves $94.75$ AUC score (with $d=4000$) which is close to that with the adjacency matrix ($95.42$ AUC). This shows that our compression retains many of the useful information in its embedding. Notice that the uncompressed embedding, for {\tt Enron}, is a 36,692-dimensional binary vector, which is almost 9 times the size of that with {\tt QUINT} with $d=4000$ and can become embarrassingly large for large sparse networks. Compression also improves the training time, e.g., training during link prediction runs out of memory on an uncompressed embedding of {\tt Gowalla}.

\noindent{\large \textbf{Organisation:}} We formally state our problem and briefly discuss our earlier work, $\binsketch$, in this context in Section~\ref{sec:problem}. Related works are discussed in Section~\ref{sec:related_work}.
Our \quint embedding method is presented in Section~\ref{sec:Background} whose theoretical properties and a few related algorithms are discussed in Section~\ref{sec:analysis}. In Section~\ref{sec:experiment}, we present the results of our empirical evaluation on several real-world datasets against  state-of-the-art algorithms. Section \ref{sec:Conclusion} concludes the paper with several open questions. \\

\noindent{\large\textbf{Reproducibility:}} The datasets that were used are described along with the specific experiments and are publicly available. We have made available the source codes of the techniques that we used, including that of \quint, on an accessible server~\cite{source_code}.

\section{Problem statement and background}\label{sec:problem}
Let $G = \langle V, E \rangle$ be an unweighted undirected network with $|V|=n$  nodes and $|E|=e$ edges. The network can also be represented  by the adjacency matrix
$ A\in \{0,1\}^{n\times n}$, where $A_{i,j}=1$ implies that there is an edge between node $i$ and node $j$, and is set to $0$ otherwise.  We use $A_i$ to denote the $i$-th row of $A$, which in turn represents a $|V|$-dimensional binary vector for the node $i$. The sparsity of $A_i$ indicates the number of ones in it, representing the degree of that node, and the sparsity of a network, denoted $\psi$, is defined as the largest sparsity of any $A_i$.\\

\noindent{{\bf Node embedding problem:}} Our aim is to compute a {\em binary embedding} of the nodes of a network, essentially an $n \times d$ binary matrix where $d$ (typically $d\ll n$) is the dimension of the embedding, in a manner that preserves local network structures.
These structures can be categorized into first, second, third-order proximities and so on, based on  their coverage~\cite{representation_learning_survey}.

{\it First-order proximity.} The first-order proximity between nodes $i$ and $j$, denoted by $s^{(1)}_{ij}$, captures the presence of an edge between them and is set to $A_{ij}$ for unweighted networks.

{\it Second and higher order proximities.} The $k$-th order proximity is defined inductively.
Let $s^{(k)}_{i}= [s^{(k)}_{i1}, s^{(k)}_{i2}, \ldots, s^{(k)}_{in} ]$ denote the $k$-th order proximity between node $i$ and the other nodes. Then the $(k+1)$-th order proximity $s^{(k+1)}_{ij}$ between nodes $i$ and $j$ is computed using a suitable ``similarity function'' between $s^{(k)}_{i}$ and $s^{(k)}_{j}$. A common practice, for example, is to use the inner product for computing similarity..
%
It should be noted that a high (or low) $k$-th order proximity between two nodes may not automatically establish a high (or low) $(k+1)$-th order proximity between them. Therefore,  different embedding techniques target proximities of specific orders~\cite{representation_learning_survey}. However, we shall explain later how {\tt QUINT}, in some manner, ensures proximities of multiple orders.\\

\noindent{\large{\bf Extension to Binsketch:}}
Recently, we proposed a randomized sketching technique for binary data, named $\binsketch$~\cite{binsketch}. It compresses high-dimensional sparse binary vectors to extremely compact binary sketches. The sketch of a vector $a \in \{0,1\}^n$ is a $d$-dimensional binary vector $\sigma_a$ whose $i$-th bit is set to $\displaystyle\bigvee_{j:\pi(j)=i} a_j$, where $\pi$ is a random mapping from $\{1,2\ldots, n\} \to \{1,2,\ldots, d\}$. We also proved  that these sketches allow efficient estimation of inner product similarity and a few other similarity metrics.

\begin{theorem}[Theorem~1,\cite{binsketch}]\label{thm:oldbs}
The inner product of two binary vectors with sparsity at most $\psi$ can be estimated from their sketches with probability at least ($1-\rho$) and accuracy $O\left(\sqrt{\tb\ln \tfrac{6}{\rho}}\right)$ if we set the size of the sketches to $d=\tb \sqrt{\tfrac{\tb}{2} \ln \tfrac{2}{\rho} }$.
\end{theorem}

The motivation behind \quint is the observation that the rows of the adjacency matrix of a network could be embedded in the above manner to low-dimension binary vectors. Further, since the inner product between two adjacency vectors represents the number of common neighbors of the corresponding nodes, it is possible to estimate the number of common neighbors between any two nodes from their embedding vectors. However, it is not clear whether this idea is practical, i.e., whether it would stand up for the common uses of node embedding in different downstream tasks, or even theoretically sound. The contribution of the current paper is an emphatic `yes' to this question, along with the modifications necessary towards this goal.

$\binsketch$ consists of two major players -- an algorithm to generate a sketch and a suite of estimators for several distance measures. The estimator for inner product distance was deemed suitable for node embedding.
Our first modification to \quint is to set a larger embedding dimension in Algorithm~\ref{algo:QUINT} for generating a sketch. The second modification is an added check in Algorithm~\ref{algo:bs} for estimating the number of common neighbors from embeddings, which turns out to be equivalent to the inner product similarity. Though a much smaller dimension appears to be effective in practice, the larger dimension, along with the check, is required to prove the theoretical guarantees on the embeddings offered by \quint, in particular Theorem~\ref{thm:binsketch} and Theorem~\ref{theorem:a4-approx}.

In our earlier work on $\binsketch$, we were solely concerned with sparse binary vectors and estimation of similarity values. Now, with all the modifications stated above, in this work we are interested in the performance of $\binsketch$ sketches when the binary vectors come from the adjacency vectors of a network. For example, we prove that when our neighborhood estimation algorithm is used, various structural properties of the network can be determined solely from the embeddings. We further report results of network analytics, namely node classification and link prediction,  on real networks.

\section{Related Work}\label{sec:related_work}
Popular node embedding methods can be broadly classified into  four categories: (1) data independent hashing, (2) learned hashing, (3) factorization based approaches, and 
(4) deep learning based approaches. 
Our method falls in the first category which creates small randomized sketches (also known as signatures in literature) without any ``learning'' from the data. The sketches are designed to allow fast computation of similarity in a suitable space. Lack of any learning step makes these efficient in time; however, the challenge is to ensure that proximities, especially those of the higher orders, can also be computed apart from the similarities. {\tt NodeSketch}~\cite{nodesketch} addresses this problem by using ``consistent weighted sampling'' that can be used to estimate a weighted variant of Jaccard similarity. Our technique, at its core, uses BinSketch ~\cite{binsketch} which can be used to estimate Jaccard, Hamming, and a few other similarity measures at the same time.

The learned hashing techniques, on the other hand, create randomized embeddings using a data dependent hash function that is learned from the data. {\tt Spectral Hashing}~\cite{SpectralHash} uses Laplacian of the corresponding graph, whereas, {\tt Scalable Graph Hashing (SGH)}~\cite{sgh} avoids computation of the entire node-to-node similarity matrix by doing a feature transformation followed by a sequential learning step.


Factorization based methods represent graph property, e.g. pairwise similarity of nodes, in the form of a matrix and factorize this matrix to obtain node embedding~\cite{GoyalF18,LianKDD2018}. {\tt GraRep}~\cite{GraRep2015} factorizes the $k$-th order transition matrix, {\tt HOPE}~\cite{HOPE_KDD} \hl{and {\tt LQANR}~\cite{lqanr}} factorize up to-$k$-order proximity matrix, and {\tt NetMF}~\cite{netmf} factorizes a matrix obtained using random walks. \hl{Recently, {\tt INH-MF}~\cite{INH-MF} was proposed to specifically learn a binary embedding using matrix factorization.} These techniques come with elegant theoretical promises, but suffers from high complexity in practice due to the optimization steps, e.g., during factorization~\cite{DemmelDH07,CaiZC18,gnn}.


Deep learning based approaches for graph embedding broadly fall into two categories. In the first category,  a graph is explored using different exploration methods, and then the contexts of the nodes obtained from the random walks are fed into the \texttt{skip-gram} model~\cite{skipgram}. \texttt{Deepwalk} preserves higher-order proximity between nodes. It uses \textit{depth-first search} to generate random walks. \texttt{node2vec} was later proposed based on the same idea, but with the difference that it employs biased random walks that provide a trade-off between \textit{breadth-first search} and \textit{depth-first search} exploration strategies. Choosing the right balance between these two enables \texttt{node2vec} to preserve \textit{community structure} and \textit{structural equivalence} between nodes. \texttt{metapath2vec}~\cite{DongCS17} extends a similar approach for Heterogeneous Networks by performing  \textit{meta-path} based random walks to construct the heterogeneous neighborhood of a node and then leverages a heterogeneous skip-gram model to perform node embeddings. 
A major limitation of such methods is that it only considers a local context within a path. Moreover, it is difficult to determine the optimal sampling strategy. The second category of algorithm does not use any random walk; instead, they model variable-sized subgraph structures in homogeneous graphs. Some examples are \texttt{GCN}~\cite{GCN}, \texttt{GraphSAGE}~\cite{Graphsage}, and \texttt{struc2vec}~\cite{struct2vec}.  To summarise, the advantage of deep learning based methods is that they are effective and robust and don’t require feature engineering, but the downside is that they incur a high computational cost.

Slightly less popular, but relevant nonetheless, are matrix-based methods. 
They are usually difficult to scale without intricate optimizations. For example, {\tt RandNE}~\cite{Zhang2018BillionScaleNE} applies SVD to billion-scale networks for which they use Gaussian random projection and the Johnson-Lindenstrauss (JL) Lemma~\cite{johnson1984extensions}. 
It should be noted that the JL Lemma is applicable primarily to real vectors and for Euclidean distance; however, we work on binary vectors, and hence, require similarity metrics relevant for binary vectors. To take advantage of bitwise operations that are known for speed and space efficiency, we use our own algorithm, {\tt BinSketch}~\cite{binsketch}.

{  Recently, sketching-based techniques have been explored for generating   node embeddings of large networks. Chen \textit{et al.}~\cite{ChenSTCS19} proposed an algorithm for computing node embedding of large   networks using very sparse random projections~\cite{LiHC06}. \texttt{FREDE}~\cite{TsitsulinMMKOM21} generates linear space embedding of nodes using deterministic matrix sketching~\cite{GhashamiLPW16}, and \texttt{InstantEmbedding}~\cite{instantembedding} computes node embedding using local PageRank computation. However, all these results lead to real-valued embedding in contrast to the space efficient binary embedding obtained via our approach.}

It is common for networks to have attributes associated with their nodes or edges or both. Several solutions have been proposed for embedding such networks. For example, 
\texttt{BANE}~\cite{bane} proposes binary embedding of attributed network by joint representation learning of node links and attributes. Along the similar lines, \texttt{ASNE}~\cite{ASNE} too generates embedding that tries to retain both structural and attribute proximities, \hl{and \texttt{GNN}~\cite{gnn} uses a graph neural network}. \texttt{DNE} (Discrete Network Embedding)~\cite{dne} is a supervised technique which learns binary node representations to speed up node classification by jointly learning the discrete embedding and classifier within a unified framework. While these are all supervised techniques, an unsupervised technique named {\tt NetHash} was proposed recently that uses a recursive hashing approach for embedding an attributed network~\cite{nethash}. In contrast, \quint is designed to embed nodes without any attributes.

{  The ``Feature hashing'' algorithm due to Weinberger \textit{et al.}~\cite{feature_hashing} takes high-dimensional real-valued data as input and outputs low-dimensional {\em real-valued} vectors (sketch) which closely approximates the original pairwise inner product similarities. If we naively apply Feature hashing on our adjacency matrix, the sketch may not remain binary \hl{({\tt node2hash}~\cite{node2hash} builds upon this idea)}. BinSketch offers a similar guarantee but generates binary sketches. In this manner, BinSketch can be seen as Feature hashing for binary data that outputs  binary vectors and allows approximation of the inner product similarity, the cosine similarity, the Jaccard similarity, and the Hamming distance.}

Many of the above techniques generate real-valued embeddings, whereas {\tt QUINT} generates binary embeddings. There are a few more ways in which our approach follows a quite different trajectory compared to the above approaches. First, we avoid direct operations on the adjacency matrix which can be extremely huge. Secondly, the above approaches try to {\em learn} the optimal embedding that minimizes a difference function; the difference function captures the desideratum of accurately obtaining some graph property (e.g., neighborhood of a node) or the effectiveness of the embedding for a certain machine learning task. {\tt QUINT} deviates from the norm and generates a {\em random} embedding upfront in an unsupervised manner. It leaves the heavy duty of using it effectively to algorithms for estimating graph properties, e.g., the degree of a node or the number of length-2 paths between two nodes.
We feel that the key factor behind the effectiveness of \quint is the effective use of the low-sparsity nature of the networks that it was designed for. None of the above techniques are explicitly designed for such networks.





 \section{\quint embedding}\label{sec:Background}

\quint uses the $\binsketch$ sketching algorithm that is appropriately designed for node embedding and modified to obtain theoretical guarantees. We describe it in Algorithm~\ref{algo:QUINT}. It uses a random mapping, denoted $\pi$, that maps every vertex to a uniformly chosen bin among $d$ bins. To embed a node $v$, it puts the endpoints of all the edges from $v$ into a uniformly and identically chosen bin using $\pi$. The embedding of $v$ is a binary vector indicating which of the $d$ bins are non-empty.

\underline{{\bf First enhancement to $\mathbf{\binsketch}$:}}
The first modification that we make is a larger embedding dimension; we set $d= \psi^2\sqrt{\frac{\psi}{2} \ln \frac{2}{\rho}}$, where $\psi$ is an upper bound on the degree of any node. This enables us to show that the sketches preserve the higher order proximities with reasonable accuracy (see Theorem~\ref{theorem:a4-approx}).


    \begin{algorithm}
    \begin{flushleft}
	\hspace*{\algorithmicindent} \textbf{Input:} undirected unweighted graph $G=\langle V,E \rangle$
	\hspace*{\algorithmicindent} \textbf{Parameter:} probability of error $\rho$
	\end{flushleft}
	\begin{algorithmic}[1]
	    \State Compute $\psi \leftarrow$ upper bound on degree of $G$
        \State Embedding dimension $d \leftarrow \psi^2\sqrt{\frac{\psi}{2} \ln \frac{2}{\rho}}$ \Comment{Increased from $\binsketch$}
	    \State $\pi \leftarrow$ a random mapping from $\{1,  \ldots, |V|\}$ to $\{1, \ldots, d\}$
	    \State For all $i=1 \ldots |V|$, initialize $\sigma_i$ as the vector $\overbrace{00\ldots 0}^d$
	    \State Set $L \leftarrow $ list of edges $E$ \label{line:QUINT1}
	    \ForAll{edge $(i,k) \in L$}
	        \State Determine $j=\pi(k)$
	        \State Set the $j$-th bit of $\sigma_i$ to 1
	    \EndFor
	    \State \Return embeddings $\{ \sigma_1, \sigma_2 \ldots \}$ of all nodes
	\end{algorithmic}
	\caption{\texttt{QUINT} embedding all nodes of a graph\label{algo:QUINT}}
    \end{algorithm}

\texttt{QUINT} falls under the ``direct encoding approaches'' in a recent categorization of graph representation methods~\cite{HamiltonYL17}. The ``encoding'' of a node $i$, represented by its adjacency vector $A_i$, is given by the matrix-vector product $P \cdot A_i^T$ where $P$ is a $d \times n$ matrix.
For \texttt{QUINT}, we choose $P$ as a random sparse binary matrix with $n$ ones such that there is exactly one $1$ in any column. For the matrix product, we use a Boolean-logic equivalent of standard matrix product, i.e., the $i$-th entry of the product of a binary matrix $M$, and a binary vector is computed as $\bigvee_{k=1}^n (M_{ik} \wedge v_k)$. An example illustrating this notion is given in Figure~\ref{fig:QUINT-matrix}.

\begin{figure}
\[
\pi:\left[
    \begin{array}{cc}
    1 \to 1 & 2 \to 3\\
    3 \to 2 & 4 \to 2\\
    5 \to 3 & 6 \to 2
    \end{array}
    \right]
    \qquad
    \left[
    \begin{array}{c}
    100000\\
    001101\\
    010010
    \end{array}
    \right] \cdot
    [010110]^T = [011]^T
\]
\caption{\label{fig:QUINT-matrix} \texttt{QUINT} embedding of a node with edges to $2,4,5$ is 011. The mapping $\pi$ is shown on the left.}
\end{figure}


 \section{Analysing \texttt{QUINT}}\label{sec:analysis}
 
\newcommand{\bs}{{\tt EstCN}\xspace}
\newcommand{\IPS}[1]{\langle #1 \rangle}
\newcommand{\hatn}{\hat{n}}
\newcommand{\hatd}{\hat{\delta}}

In this section, we analyse the theoretical properties of \quint that make it suitable for embedding graphs while retaining information about its structural properties, and this, we believe, helps in node classification and link prediction. It will be helpful to remember that the embedding of a node $i$ is a $d$-dimensional binary vector which is denoted $\sigma_i$ and is defined as:\\
\centerline{
{\bf $\mbox{ $j$-th coordinate of $\sigma_i$: } (\sigma_i)_j  = \bigvee_{k: \pi(k)=j} A_{ik}$}}
where $\pi$ denotes a random mapping from $\{1, 2, 3, \ldots, n\}$ to $\{1, 2, 3, \ldots, d\}$ and $A$ is the adjacency matrix.

 
 
\subsection{Time and space complexity}

The random mapping can be generated once and stored as a fast-lookup table or can be implemented using an efficient hashing algorithm. Let $S$ denote the space required and $T$ denote the time-complexity to compute $\pi(\cdot)$; for a lookup table stored in RAM, $S = O(n \log d)$ and $T = O(1)$. Apart from the overhead of $\pi$, \texttt{QUINT} only involves bit manipulations which makes it extremely efficient. We assume that $G$ is stored as an adjacency list (which has asymptotically same space complexity as that of an edge-list representation).

\begin{lemma}
Algorithm~\ref{algo:QUINT} runs in time $\Theta(T \cdot |E|)$, returns an embedding using $d \cdot |V|$ bits and requires an additional space $S$ for the embedding.
\end{lemma}

We point out that $\pi$ does not need to be stored once a graph is embedded. While the node embeddings can be stored in RAM, another alternative is to not store the embedding, but to store only $\pi$ and generate $\sigma_i$ on demand; this approach is also efficient given the lightweight embedding algorithm.

\begin{lemma}
 The embedding of a single node  $i$ uses only $d$ bits and can be computed in time $O(T \cdot n(i))$, where $n(i)$ denotes the degree of $i$, and without any additional space overhead.
\end{lemma}

Taking $T$ and $d$ to be constant for practical scenarios, both these approaches take constant time per edge of the network and use constant space per node of the network.



\subsection{Estimating number of neighbors}
The {\em number of neighbors} of a node can be easily estimated from its embedding by re-using a result that was proved by us for $\binsketch$ except that we use a larger $d$ compared to what was used there. Choosing a larger reduced dimension obviously does not adversely affect the accuracy.

    
    
\begin{lemma}[Lemma~8,\cite{binsketch}]\label{lem:degree}
There is a linear-pass algorithm to estimate the number of neighbors of a node $i$ with at most $4\sqrt{\frac{\psi}{2}\ln \frac{2}{\delta}}$ additive inaccuracy and probability at least $1-\delta$.
\end{lemma}

\subsection{Estimating lower order proximities}
We used the ``inner product'' similarity measure for node classification and link prediction. The first-order proximity between two nodes, say $s_{ij}^{(1)}$, is always the presence of an edge between them, i.e., same as $A_{ij}$. To compute the second-order proximity, observe that $s_i^{(1)}=[s_{i1}^{(1)}, s_{i2}^{(1)}, \ldots s_{in}^{(1)}]$ which is identical to $A_i$. Now, $s_{ij}^{(2)}$ is the inner product of $s_i^{(1)}$ and $s_j^{(1)}$ and this is same as the inner product of $A_i$ and $A_j$. Further, the latter can be easily shown to be the number of common neighbors of $i$ and $j$ --- denoted $n_{i,j}$ (this can also be represented as the $(i,j)$-th entry of $A^2$). Note that $n_{i,j} \le \psi$, degree of the graph. 

First we show that \quint allows us to estimate the first-order proximity for the  inner product similarity measure. The first-order proximity between nodes $i$ and $j$ indicates the presence of an edge between them. There was no mention of anything similar in $\binsketch$ so we show how this can be estimated from their sketches $\sigma(i)$ and $\sigma(j)$ using Algorithm~\ref{algo:first-order}.

    \begin{algorithm}
    \begin{flushleft}
    \hspace*{\algorithmicindent} \textbf{Input:} Nodes $i$ and $j$\\
	\hspace*{\algorithmicindent} \textbf{Input:} Embeddings $\sigma_i$ of $i$ and $\sigma_j$ of $j$\\
	\end{flushleft}
	\begin{algorithmic}[1]
	    \State Compute $a = \pi(i)$
	    \State Compute $b = \pi(j)$ \Comment{$a,b$ could be same or different}
	    \If{$b$-th bit of $\sigma_i$ and $a$-th bit of $\sigma_j$ are both 1}
	        \State\Return {\tt true}
	    \Else
	        \State \Return {\tt false}
	    \EndIf
	\end{algorithmic}
	\caption{Estimate presence of an edge between nodes $i,j$\label{algo:first-order}}
    \end{algorithm}

It is obvious that if there is an edge between nodes $i$ and $j$ then Algorithm~\ref{algo:first-order} is correct, since $i_j=j_i=1$ and that implies that the $b$-th bit of $\sigma_i$ is 1 and so is the $a$-th bit of $\sigma_j$. The other case of no edge between $i$ and $j$ is covered by Lemma~\ref{lem:first-order}.

\begin{lemma}\label{lem:first-order}
Algorithm~\ref{algo:first-order} is always correct when there is an edge between $i$ and $j$ and has a probability of error at most $\tfrac{2\psi}{d} \le \tfrac{2\sqrt{2}}{\psi\sqrt{\psi \ln 2/\rho}}$ when there is no such edge.
\end{lemma}

\begin{proof}
The proof of the first case is given above. The proof of the second case is obtained by bounding the probability of events $E_i=``\pi(k) \not=\pi(j)$ for all neighbor $k$ of $i"$ (there are at most $\psi$ neighbors) and $E_j$ is defined similarly for $j$. Note that the probability that the estimator is correct can be expressed as $\Pr[E_i \cup E_j]$ which is at least $\Pr[E_i] + \Pr[E_j] - 1$. The proof follows from the observation that $\Pr[E_i] \ge (1-\tfrac{1}{d})^{\psi}$.
\end{proof}

Next we show that \quint\ allows us to estimate the second-order proximity using the above similarity measure -- the number of common neighbors. We describe this approach in Algorithm~\ref{algo:bs}. Given embeddings $\sigma_i$ and $\sigma_j$ of nodes $i$ and $j$, respectively, the algorithm first determines the number of ones in those embeddings. Observe that the number of ones in the $i$-th row of $A$ determines the degree of $i$. Conceptually the algorithm will use the number of ones in $\sigma_i$ to determine the number of ones in $A_i$, albeit with some inaccuracy, and thereby estimate its degree.

Then the algorithm computes the number of common indices where both $\sigma_i$ and $\sigma_j$ are one, denoted $\hatn^s_{i,j}$. Once again, the number of common indices where both $A_i$ and $A_j$ are one is exactly the number of common neighbors they have, and, will be estimated from $\hatn^s_{i,j}$. It turns out this estimation requires the degrees of $i$ and $j$ whose approximate values we calculated above.


    \begin{algorithm}
    \begin{flushleft}
    \hspace*{\algorithmicindent} \textbf{Input:} Nodes $i$ and $j$\\
	\hspace*{\algorithmicindent} \textbf{Input:} Embeddings $\sigma_i$ of $i$ and $\sigma_j$ of $j$\\
	\hspace*{\algorithmicindent} \textbf{Parameter:} Embedding dimension $d$
	\end{flushleft}
	\begin{algorithmic}[1]
	    \State Compute $\hatn^s_{i}=|\sigma_i|$ \Comment{$|v|$: number of ones in vector $v$}
	    \State Compute $\hatn^s_j=|\sigma_j|$
	    \State Compute $\hatn^s_{i,j}=\IPS{\sigma_i, \sigma_j}$ \Comment{$\IPS{u,v}$: inner prod. of $u$ and $v$}
	    \If{$\hatn^s_{i,j}=0$}
	        \State \Return $\hatn_{i,j}=0$ \label{line:algobs-sp-case} \Comment{Enhancement from $\binsketch$}
	    \EndIf
	    \State Set $\hatn_i=\ln(1-\frac{\hatn^s_i}{d})/\ln(D)$ \Comment{$D$ denotes $1-\frac{1}{d}$}
	    \State Set $\hatn_j =\ln(1-\frac{\hatn^s_j}{d})/\ln(D)$
	    \State \Return estimated number of common neighbors $$\hatn_{i,j} = \hatn_i + \hatn_j - \tfrac{1}{\ln D} \ln
		\left(D^{\hatn_i}+D^{\hatn_j}+\frac{\hatn^s_{i,j}}{d}-1 \right)$$ 
	\end{algorithmic}
	\caption{\bs (number of common neighbors $n_{i,j}$)\label{algo:bs}}
    \end{algorithm}


The next theorem formalizes our claim that the embeddings preserve the second order proximities.

\begin{theorem}\label{thm:binsketch}
Let $\psi$ be an upper bound on the degree of $G$ and $\rho$ be the desired probability of error. If the embedding dimension $d$ is chosen as $\psi^2 \sqrt{\frac{\psi}{2} \ln \frac{2}{\rho}}$, then \bs ensures that its output satisfies the following with probability $\ge 1-\rho$.
$$(1) \qquad n_{i,j} - 14\sqrt{\frac{\psi}{2} \ln \frac{6}{\rho}} \quad  <   \hatn_{i,j}  < \quad n_{i,j} + 14\sqrt{\frac{\psi}{2} \ln \frac{6}{\rho}}$$
{\em (2)} Furthermore, if $n_{i,j} > 0$ then $\hatn_{i,j} > 0$, and if $n_{i,j}=0$ then $\hatn_{i,j}=0$ with probability at least $1- \sqrt{2/(\psi \ln \tfrac{2}{\rho})}$.
\end{theorem}

\underline{\bf Second enhancement to $\mathbf{\binsketch}$:} We quickly discuss the role of the enhancements to $\binsketch$ in proving this theorem. The first part about the accuracy of $\bs$ in Theorem~\ref{thm:binsketch} is known to be a feature of the \texttt{BinSketch} algorithm (refer to Theorem~\ref{thm:oldbs}), except that the embedding dimension used there was $\psi \sqrt{\frac{\psi}{2} \ln \frac{2}{\rho}}$. Our first modification was a larger value $\psi^2\sqrt{\frac{\psi}{2} \ln \frac{2}{\rho}}$. But since the embedding dimension has increased, there are now even lesser chances of collision in the computation of $\pi(\cdot)$ and higher chances that a particular bit of an embedding is set by {\em only one bit} of an input vector. Intuitively speaking, this will only result in a better accuracy and lower probability of error during an estimation compared to the that by the \texttt{BinSketch} algorithm.

The second part says that, with high probability, $n_{i,j}=0$ iff $\hatn_{i,j}=0$. This is important, since otherwise, the {\em estimated} number of common neighbors of $i$ and $j$ could be as large as $14\sqrt{\frac{\psi}{2} \ln \frac{6}{\rho}}$ even when $i$ and $j$ share no common neighbor.  The enhancement in Line~\ref{line:algobs-sp-case} of Algorithm~\ref{algo:bs} is crucial to prove the second part. 

Both the modifications in Algorithm~\ref{algo:QUINT} and Algorithm~\ref{algo:bs} are neither cosmetic nor heuristic but are necessary to prove Theorem~\ref{thm:binsketch}, especially that $\hatn_{i,j}$ is concentrated around $n_{ij}$ in the second part. This, in turn, enables us to prove Theorem~\ref{theorem:a4-approx} which says that certain higher order proximities are preserved by \quint.

\begin{proof}[Proof sketch for the first part] We present a sketch of the proof here; for the detailed calculations, please refer to our earlier work on $\binsketch$~\cite{binsketch}.

Due to the uniform nature of the random mapping $\pi$ used in Algorithm~\ref{algo:QUINT}, it is easy to show that $\E[\hatn^s_{i}]=d\big( 1-D^{deg(i)} \big)$ in which $deg(i)$ denotes the degree of the node $i$; a similar identity also holds for $\hatn^s_{j}$. These formul\ae\ allow us to express $deg(i)$ in terms of $\E[\hatn^s_i]$.

Next, observe that the $t$-th bit of both $\sigma_i$ and $\sigma_j$ can be set only in one of two ways.
\begin{enumerate}
    \item Both $i$ and $j$ have an edge to some node $k$ which is mapped by $\pi$ to $t$.
    \item Nodes $i$ and $j$ do not share a common neighbor but the $t$-th bit is set due to some neighbor $a$ of $i$ for which $\pi(a)=t$ and due to some neighbor $b$ of $j$ for which $\pi(b)=t$.
\end{enumerate}
Once again the uniform nature of $\pi$ allows us to express the expected number of bits that are set in both $\sigma_i$ and $\sigma_j$ by the following formula.\\
\centerline{$\displaystyle \E[\hatn^s_{i,j}] = d \Big( 1 - D^{deg(i)} - D^{deg(j)} + D^{deg(i) + deg(j) + n_{i,j}} \Big)$}\\

This expression allows us to express $n_{i,j}$ in terms of $\E[\hatn^s_{i,j}]$, $deg(i)$ and $deg(j)$, and further, in terms of $\E[\hatn^s_{i,j}]$, $\E[\hatn^s_i]$ and $\E[\hatn^s_j]$ using the two formul\ae\ given above. This is precisely what Algorithm~\ref{algo:bs} does. Of course, it does not have the actual expectation values. But fortunately we can show that the observed values of $\hatn^s_{i,j}$, $\hatn^s_i$ and $\hatn^s_j$ are tightly concentrated around their expectations. In other words, it suffices to use the observed values in place of their expectations in lieu of some inaccuracy in $\hatn({i,j})$; the inaccuracy too can be bounded by carefully combining the concentration bounds of $\hatn^s_{i,j}$, $\hatn^s_i$ and $\hatn^s_j$.
\end{proof}

\begin{proof}[Proof of second part]
For the first claim, we observe that if $n_{i,j} > 0$, then there must be some $k$ such that $A_{ik}=A_{jk}=1$. Let $t$ denote $\pi(k)$; then both $(\sigma_i)_t=(\sigma_j)_t$ will be set to 1. Thus, $\hatn^s_{i,j} = \IPS{\sigma_i,\sigma_j} > 0$. By way of contradiction, assume that $\hatn_{i,j} = 0$. But then we can derive the following identities based on Algorithm~\ref{algo:bs}.

\begin{align*}
  \hatn_i + \hatn_j & = \tfrac{1}{\ln D} \ln
		\left(D^{\hatn_i}+D^{\hatn_j}+\frac{\hatn^s_{i,j}}{d}-1 \right)\\
	  D^{\hatn_i} \cdot D^{\hatn_j} & = D^{\hatn_i} + D^{\hatn_j} + \frac{\hatn^s_{i,j}}{d} - 1 \\
	  \left(1 - \frac{\hatn^s_i}{d} \right) \left(1 - \frac{\hatn^s_j}{d} \right) & = \left(1 - \frac{\hatn^s_i}{d} \right) + \left(1 - \frac{\hatn^s_j}{d} \right) + \frac{\hatn^s_{i,j}}{d} - 1\\
	  \frac{\hatn^s_i \cdot \hatn^s_j}{d^2} & = \frac{\hatn^s_{i,j}}{d} \implies \frac{|\sigma_i|}{d} \cdot \frac{|\sigma_j|}{d} = \frac{\IPS{\sigma_i,\sigma_j}}{d}
\end{align*}
The last identity may not always hold. For example, consider a scenario in which $\sigma_i \sim \sigma_j$ with the number of ones in each much less than $d$; in this case $|\sigma_i| \approx |\sigma_j| \approx \IPS{\sigma_i, \sigma_j}$ and $|\sigma_i| \ll d$, thus contradicting the identity. This proves that $\hatn_{i,j} > 0$.

For proving the second claim, take any $i$ and $j$ such that $n_{i,j}=0$, i.e., there is no $k$ such that both $A_{i,k}=A_{j,k}=1$. For the sake of contradiction, assume that $\hatn_{i,j} > 0$, i.e.,  there is some $x$ such that $(\sigma_i)_x = (\sigma_j)_x = 1$; this means that there must be some $k_1$ and $k_2$ such that $A_{i,k_1}=A_{j,k_2} = 1$ and $\pi(k_1)=\pi(k_2)=x$. The probability for the latter event, denoted $E$, is same as the probability that in a group of $|A_i|$ men and $|A_j|$ women, there exists at least one pair with a common birthday, which can be shown to be $$1 - \sum_{k=1}^{|A_i|} {D \choose k} k! \, S_{|A_i|,k} (d-k)^{|A_j|}/d^{|A_i|+|A_j|}$$
in which $S_{\cdot,\cdot}$ denotes Stirling's number of the second kind~\cite{stackxch-birthday}. Getting a closed form of this is difficult; therefore, we show a different technique of bounding the probability of $E$.

Let $E_x$ be the event that $(\sigma_i)_x= (\sigma_j)_x=1$. It can be shown that $\E(E_x) =  \left(1-D^{|A_i|}\right) \cdot \left(1-D^{|A_j|}\right)$~\cite[Lemma 5]{binsketch}. Since $|A_i|$ and $|A_j|$ are both at most $\psi$, and $D < 1$, 
$$\E(E_x) \le (1-D^{\psi})^2 = \left(1 - \left(1-\frac{1}{d}\right)^{\psi}\right)^2 \le \left(\frac{\psi}{d}\right)^2$$
which implies that $\E[\sum_x E_x] = d\left(\frac{\psi}{d}\right)^2 = \frac{\sqrt{2}}{\sqrt{\psi\ln \frac{2}{\rho}}} \ll 1$

Let $E_{all}$ denote $\sum_x E_x$; $E_{all}$ is a random variable that denotes the number of positions $x$ such that $(\sigma_i)_x=(\sigma_j)_x=1$ and whose expectation we computed above. Since $E$ is equivalent to ``$E_{all} \ge 1$'', we apply Markov's inequality to bound it.
\end{proof}

For the rest of this section, we will use the fact that \bs estimates non-zero $n_{i,j}$ values with a small additive error and accurately identifies $n_{i,j}=0$ values (both of these happen with non-negligible probability which we would not explicitly state to simplify calculations).


\renewcommand{\L}{\mathcal{L}}

The above theorem allows us to accurately quantify the loss function ({\it aka.} objective function) that is used by most node embedding algorithms to {\em learn} the embedding~\cite{HamiltonYL17}. Using the mean-squared-error (MSE) to compute the loss function, we can represent it as
$$\L = \frac{1}{T}\sum_{(i,j) \in T \subseteq V \times V} | n_{i,j} - \bs(i,j) |^2.$$
Here $T$ represents a ``training set of edges'' that is traditionally used to learn an embedding. Our proposed technique does not involve any learning; nevertheless, for the sake of comparison we present an explicit upper bound on $\L$ --- it follows directly from Theorem~\ref{thm:binsketch}.

\begin{lemma}
Using \texttt{QUINT} for embedding and \bs for estimating node similarity, $\L$ is upper bounded by $196 \frac{\psi}{2} \ln \frac{6}{\rho}$.
\end{lemma}

\subsection{Estimating higher-order proximities}

Many node embedding algorithms operate on the paths in a graph, often up to a certain length. For example, both the random-walk based approaches, \texttt{deepwalk} and \texttt{node2vec}, consider two nodes to be similar if their presence in short random walks on the graph are highly correlated. Here we show that \texttt{QUINT} also ``preserves'' path information in a certain manner. The key observation here is that the square of the adjacency matrix precisely contains the $n_{ij}$ values.

\begin{obs}
For any $i,j$, $A^2_{i,j} = n_{i,j}$.
\end{obs}
This holds due to the following calculation. $$\sum\limits_{k = 1}^{n} A_{ik} A_{kj} = | \{ k \in V ~:~ (i, k) \in E,~ (k, j) \in E\}|.$$

This observation along with Theorem~\ref{thm:binsketch} implies that \bs can approximately compute $A^2$.

\begin{lemma}\label{lemma:approx_A2}
$\bs(\sigma_i,\sigma_j)$ approximately computes $A^2_{i,j}$ with a small additive error. If $A^2_{i,j}=0$ $\bs(\sigma_i,\sigma_j)$ outputs 0.
\end{lemma}

Next we show how to translate the above lemma to higher {\em even} powers of $A$. We first show the result for the $4$-th power, $A^4$, each of whose entry, say $A^4_{i,j}$, denotes the number of paths between $i$ and $j$ using 4 nodes (and 3 edges).

We use $\epsilon_2$ to denote the small additive error mentioned in Lemma~\ref{lemma:approx_A2}. To maintain consistency of notations, we will use $\hatn2_{i,j}$ to denote $\hatn_{i,j}$; by construction, $\hatn2$ is symmetric.

\begin{theorem}\label{theorem:a4-approx}
Let $\hatn4_{i,j}$ denote the expression $\sum_{k=1}^n \hatn2_{i,k} \cdot \hatn2_{k,j}$ for all $i,j=1 \ldots n$. Then
$\hatn4_{i,j}$ approximately computes $A^4_{i,j}$ with a small additive error. If $A^4_{i,j}=0$ then $\hatn4_{i,j}$ outputs 0.
\end{theorem}

\begin{proof} Recall that $A^4_{i,j} = \sum_{k=0}^n A^2_{i,k} A^2_{k,j} = \sum_{k=0}^n A^2_{i,k} A^2_{j,k}$, and further more, each term in the summation is non-negative.
Therefore, $A^4_{i,j}=0$ implies that $A^2_{i,k}=0$ and $A^2_{j,k}=0$ for every $k=1 \ldots n$. We know from Lemma~\ref{lemma:approx_A2} that, in this case, $\hatn2_{i,k}=0$ and $\hatn2_{j,k}=0$ for all $k$. Clearly, $\hatn4^{i,j}=0$ too --- this proves the second part of the theorem.

For the first part, take any $x,y$ such that $A^4_{x,y} > 0$. That is, $\sum_{k=1}^n A^2_{x,k} A^2_{y,k} > 0$. 
From Lemma~\ref{lemma:approx_A2} we know that ${|\hatn2_{x,y} - A^2_{x,y}| \le \epsilon_2}$; using $z$ to denote the left-hand side, we can write $\hatn2_{x,y} = A^2_{x,y} + z$ where $-\epsilon_2 \le z \le \epsilon_2$.

We now state a technical result about $A$.
\begin{claim}\label{claim:1}
$\sum_{u=1}^n A^2_{x,u} \le \psi^2$.
\end{claim}
\begin{subproof}
$\psi$ being an upper bound on the degree of any node, the total number of length-2 paths from $x$ is at most $\psi^2$. $A^2_{x,u}$ is the total number of length-2 paths from $x$ to $u$ and $\sum_{u} A^2_{x,u}$ is the total number of length-2 paths from $x$ to any node, which is, therefore, upper bounded by $\psi^2$.
\end{subproof}
The next observation is applicable to sparse graphs in general but immensely beneficial to graphs where $\psi^2 \ll n$.
\begin{obs}\label{obs:2}
Since the entries of $A^2_{x,u}$ are non-negative, Claim~\ref{claim:1} implies that at most $\psi^2$ entries are non-zero in the $x$-th row of $A^2$, i.e, among $A^2_x = \{ A^2_{x,u} ~:~ u \in \{1, 2, \ldots, n\} \}$. That is, at least $n-\psi^2$ entries of $A^2_x$ are zero.
\end{obs}

This observation, along with Lemma~\ref{lemma:approx_A2}, implies that for any $x$, at most $\psi^2$ values in the set $\{\hatn2_{x,u} ~:~ u \in \{1, 2, \ldots n\} \}$ are non-zero. We use $(\sum_{u=1}^n)^{\le \psi^2}$ to denote the fact that in a summation with $n$ summands, at most $\psi^2$ terms are non-zero. Getting back to proving the theorem,
\begin{align*}
    \hatn4_{x,y}
    = &  \sum_{u=1}^{n} \hatn2_{x,u} \cdot \hatn2_{y,u} \\
    = & (\sum_{u=1}^n)^{\le \psi^2} (A^2_{x,u} + z) \cdot (A^2_{y,u} + z)\\
    = & \sum_{u=1}^n A^2_{x,u} \cdot A^2_{y,u} + z \cdot \sum_{u=1}^n A^2_{x,u} \\
    & + z \cdot \sum_{u=1}^n A^2_{y,u} + (\sum_{u=1}^n)^{\le \psi^2} z^2 \\
    = & A^4_{x,y} + 2z\psi^2 + z^2 \psi^2 \tag*{(Using Claim~\ref{claim:1})}
\end{align*}
Thus we get, $$\big|\hatn4_{x,y} - A^4_{x,y}\big| = \big|2z + z^2\big| ~ \psi^2 = \left\{
\begin{array}{ll}
    3|z|\psi^2 & \mbox{ if $|z| < 2$}\\
    2z^2 \psi^2 & \mbox{ if $|z| \ge 2$}
\end{array} \right.$$
Since $|z| \le \epsilon_2$ and the additive error $\epsilon_2$ for $\hatn^2_{x,y}$ is $\tilde{O}(\sqrt{\psi})$ from Theorem~\ref{thm:binsketch} (here $\tilde{O}()$ hides $\log(1/\rho)$ factors), so both $|z|,z^2$ is $\tilde{O}(\psi)$. Therefore, the additive error for $\hatn4_{x,y}$ is $\tilde{O}(\psi^3)$.
\end{proof}

Theorem~\ref{theorem:a4-approx} can be generalized to show that entries of $A^{2^t}$, for $t \ge 1$, can be approximated with additive error $\tilde{O}(poly(\psi))$. Since any even power of $A$ can be written as a product of $A$ raised to a power of 2, we have established that our \texttt{QUINT} embedding effectively preserves all even powers of the adjacency matrix $A$, and therefore, can approximate information about paths of even lengths in $G$. 
\section{Empirical Evaluation}\label{sec:experiment}
We first describe our experimental setup. Then we report how our proposed solution performs on two end tasks -- node classical and link prediction, with specific emphasis on speed and quality.

{\textbf{Hardware description:}} We performed most of our experiments on a laptop with the following configuration: Intel(R) Core(TM) i7-4710MQ  CPU @ 2.50GHz x 8, 7.5 GB RAM, Ubuntu 18.04 64-bits OS.
 The experiments on the \texttt{Gowalla}~\cite{snap}, {\tt Youtube}~\cite{dataset}, and {\tt Flickr}~\cite{dataset} datasets were performed on a server with the following configuration: Intel(R) Xeon(R) CPU E5-2650 v3 @ 2.30GHz, {94} GB RAM, Ubuntu 64-bits OS.

 {\textbf{Baseline methods:}}
We evaluated our approach against seven state-of-the-art node embedding methods -- \texttt{node2vec}~\cite{node2vec}, \texttt{deepwalk}~\cite{deepwalk},   \texttt{LINE}~\cite{line}, \texttt{VERSE}~\cite{verse}, \texttt{NetMF} \cite{netmf},   \texttt{NodeSketch}~\cite{nodesketch}, and \texttt{SGH}~\cite{sgh}; {\tt SGH} generates binary embedding  like {\tt QUINT}. { Since {\tt QUINT} generates the embedding of a node by compressing its adjacency vector, we also performed experiments using the uncompressed  adjacency matrix where the $i$-th row of the matrix is used as the embedding of the $i$-th node; we refer to this method as ``{\tt Uncompressed}''.}


\begin{table}
\begin{center}
\caption{Datasets used for link prediction.}
\begin{tabular}{ |c|c|c|c| } 
 \hline
 Dataset & Nodes &Edges & Max. degree \\ 
 \hline
 Gowalla & 196,591 & 950,327  & 14,730 \\
 Enron Emails Network & 36,692 & 183,831 & 1,383 \\ 
 Facebook & 4,039 & 88,234 & 1,045 \\
 BlogCatalog & 10,312 & 333,983 & 3,992 \\ 
 { Flickr}   & 80,513 &  5,899,882 & 5,706\\
 { Youtube}   & 1,138,499 & 2,990,443  & 28,754\\
 \hline
\end{tabular}
\label{tab:data_link_prediction}
\end{center}
\vspace{-5mm}
\end{table}

\begin{table}\begin{center}
\addtolength{\tabcolsep}{-3pt}
\caption{Comparison of space required for storing embeddings}
\begin{tabular}{ @{} | p{4cm} | p{1.3cm} | p{2.8cm} |@{}}
 \hline
 Method & dimension &  $\#$ of binary bits \\ 
 \hline
 \texttt{node2vec}, \texttt{deepwalk}, \texttt{LINE}, \texttt{VERSE}, \texttt{NetMF}, \texttt{NodeSketch} & 128 & $128 \times 64 = 8,192$\newline (64-bits for floating pt.) \\
 \hline
 \texttt{QUINT, SGH} & $d$ & $d$\\ 
 \hline
{  {\tt Uncompressed}} & { $|V|$} & { $|V|$ (no. of nodes)}\\
 \hline
\end{tabular}
\label{tab:bits_space}
\addtolength{\tabcolsep}{3pt}
\end{center}\end{table}

\begin{table*}
\centering
\caption{Comparison on AUC-ROC and compression time of \texttt{QUINT} and other baselines (using 128 dimensions) for the link prediction experiments.
For the datasets on which {\tt QUINT} outperformed the baselines, we have reported the  performance of {\tt QUINT} on the smallest dimension at which it outperformed the latter; for the other datasets we have reported the best performance of {\tt QUINT} (see Table~\ref{tab:link_prediction_full_128} in Supplementary for the results for all the dimensions). 
We stopped baselines that took 10 hours or more and indicate them by \DNS; {\tt OOM} indicates baselines that ran out of memory. Smaller datasets were embedded to a maximum of 4000 dimensions. 
The best AUC-ROC among the baselines and for {\tt QUINT} are {\bf bolded}.}

    \noindent \begin{tabular}{|l|ccc| ccc| ccc|}
    \toprule
    \multirow{1}{*}{Method} &
      \multicolumn{3}{c|}{Gowalla} &
      \multicolumn{3}{c|}{Flickr} &
      \multicolumn{3}{c|}{Youtube}\\
       & {Dim.} & {AUC-ROC} &{Time(s)}  & {Dim.} & {AUC-ROC} &{Time(s)}  & {Dim.} & {AUC-ROC} &{Time(s)}  \\

            \midrule
     \texttt{node2vec} & 128  & 75.48 & 7827  
                       & 128  & 76.4 & 8706.8 
                       & 128  & 65.1 & 30899.28 \\
        \texttt{deepwalk} & 128  & 74.79 & 7075
                          & 128  &  67.7    & 4978.64
                          & 128  &  \NA[55.4]    & \DNS[91783.87] \\

        \texttt{LINE}      & 128 & \NA[75.33] & \DNS[72242]
                           & 128 &  \NA[41.5]   & \DNS[186796.55]
                           & 128 &  \NA[47.69]   & \DNS[1219860.63] \\

       \texttt{VERSE} & 128 & \textbf{87.35} & 4902.8 
                      & 128 & {\bf 91.6}  & 9469.08 
                      & 128 & \NA[{\bf 87.04}]  & \DNS[174982.02]  \\
                     
         
          \texttt{NetMF} & 128 & \NA  & \OOM
                        & 128 &  \NA & \OOM
                        & 128 &  \NA  & \OOM \\

         \texttt{NodeSketch} & 128  & 50.02 & 328.30 
                             & 128  & 50.03 & 770.65
                             & 128  & 50.04 & 538.81 \\
                            
         
          \texttt{SGH} & 128 & \NA  & \OOM
                        & 128 &  \NA & \OOM
                        & 128 &  \NA  & \OOM \\

                       
         \hline
         {\tt QUINT} & 1000 & {\bf 87.71} & 7.15 & 100 & {\bf 86.7} & 7.41 & 100 & {\bf 90} & 8.86 \\




                        




                        
         \hline
         \texttt{Uncompressed}   & \NA & \texttt{OOM}   & \NA  
                                 & \NA & \texttt{OOM} & \NA  
                                 & \NA & \texttt{OOM} & \NA \\
                                 
    \bottomrule
  \end{tabular} \\[1em]

  \begin{tabular}{|l|ccc| ccc| ccc|}
    \toprule
    \multirow{1}{*}{Method} &
      \multicolumn{3}{c|}{Enron} &
      \multicolumn{3}{c|}{BlogCatalog} &
      \multicolumn{3}{c|}{Facebook} \\
         & {Dim.} & {AUC-ROC} &{Time(s)}  & {Dim.} & {AUC-ROC} &{Time(s)}  & {Dim.} & {AUC-ROC} &{Time(s)} \\

            \midrule
     \texttt{node2vec} 
                       & 128  & 67.13 & 560  
                       & 128  & 63.12 & 979.02
                       & 128  & 93.64 & 63.98\\
        \texttt{deepwalk} 
                          & 128  & 66.7  & 515.36
                          & 128  & 61.00    & 774.31
                          & 128  & 93.10 & 54.39\\
                          
        \texttt{LINE}   
                           & 128 & 75.87 & 4216
                           & 128 & 66.08    & 2379
                           & 128 & 83.41 & 250.25\\
                           
       \texttt{VERSE} 
                      & 128 & \textbf{97.20} & 1058.42 
                      & 128 & 74.24          & 351.38 
                      & 128 & 94.32          & 88.34\\
         
         \texttt{NetMF}   
                        & 128 & 83.98         & 62.29 
                        & 128 & 71.68         & 7.7 
                        & 128 & {\bf 94.86} & 2.02\\
                        
         \texttt{NodeSketch} 
                             & 128  & 49.99 & 44.07
                             & 128  & 50.00 & 75.53
                             & 128  & 50.00 & 10.69\\
         \hline
         
          \texttt{SGH} 
                       & 128 & 71.49        & 31.28
                       & 128 & 70.05        & 7.53
                       & 128 & 92.20        & 5.3\\
                       
         \texttt{}     
                       & 1600 & 68.00       &106.01
                       & 2400 & {\bf 75.36}       & 84.74
                       & 1800 & 88.89       & 50.32\\
         \hline
         {\tt QUINT} & 4000 & {\bf 94.75} & 9.05 & 100 & {\bf 81.60} & 0.88 & 4500 & {\bf 95.30} & 0.53 \\









         \hline
         \texttt{Uncompressed}     
                                 & \NA & 95.42 & \NA  
                                 & \NA & 86.46 & \hspace*{0.32cm}\NA \hspace*{0.32cm} 
                                 & \NA & 95.44 & \hspace*{0.3cm} \NA\hspace*{0.3cm}\\
    \bottomrule
  \end{tabular} \\[1em]

  \label{tab:link_prediction}
\end{table*}

For \texttt{node2vec}, we used the implementation provided by its authors\footnote{https://github.com/aditya-grover/node2vec}. We did a grid search over its parameters $p$ and $q \in [0.25, 0.5, 1, 2, 4]$, which  is equivalent to running $25$ different experiments. Here we only 
 report the result for the optimum choice of $p$ and $q$. \texttt{deepwalk} is a special case of  \texttt{node2vec} when $p=q=1$~\cite{node2vec}, so the earlier {\tt node2vec} implementation was used here with the specified parameters. For \texttt{LINE}, we used a standard  implementation available online 
\footnote{\url{https://github.com/shenweichen/GraphEmbedding}} and performed experiments considering both first and second order proximity --- and reported the best result. \texttt{VERSE}~\cite{verse} \footnote{\url{https://github.com/xgfs/verse}} uses the idea of personalised PageRank to compute the embedding of nodes. 
\texttt{NetMF}~\cite{netmf}\footnote{\url{https://github.com/xptree/NetMF}} unifies the idea of sampling based algorithms such as \texttt{node2vec, deepwalk, LINE} in the  matrix factorisation framework to generate embeddings. 
 \texttt{NodeSketch}~\cite{nodesketch}\footnote{\url{https://github.com/eXascaleInfolab/NodeSketch}} is built on top of an efficient sketching technique, and it outputs embeddings which preserve higher-order proximity via a recursive approach. 
 \texttt{SGH}~\cite{sgh}\footnote{\url{4https://github.com/jiangqy/SGH-IJCAI2015}} learns the binary embedding of nodes by minimizing the difference between the similarity of a pair of nodes and the Hamming similarity of the corresponding binary hash codes.



The embeddings obtained from \texttt{QUINT}, \texttt{SGH}, and full adjacency matrix representation are binary, whereas the other algorithms output real-valued vectors. A real-valued vector that is represented using 64-bit floating point numbers (a common configuration) takes $64\times$ more space as compared to a binary vector of the same dimension. We summarise this in Table~\ref{tab:bits_space}. For all the algorithms which give real-valued embeddings,  we {  report the results of embedding in $128$ dimensions in this section. The results for embedding in 256 dimensions are similar and are given in Tables~\ref{tab:node_classification_256}, \ref{tab:blogcatalog_node_class}, and \ref{tab:link_prediction_full_256} in Supplementary.}


\subsection{Comparing performance on link prediction}\label{subsec:link_prediction}

In the link prediction problem, we are given a network with a certain fraction of missing edges, and the task is to predict these missing edges. It can be recast as a classification problem where the goal is to train a classifier that, given a pair of nodes, outputs if there is an edge between them. 

For the purpose of classification, we generated a labeled dataset of edges by splitting a dataset into training and testing partition in $70-30$ ratio. \hl{To generate the positive training and testing samples,} we randomly sampled $30\%$ of edges and removed them. During the removal of edges, we made sure that the residual graph obtained after edge removal remains connected. If the sampled edges do not ensure this, we chose a different edge. We then generated ``missing edges'' equal in number to \hl{those in} the original dataset --- these are ``edges'' that are absent in the graph. We split the generated missing edges into training and testing partition in $70-30$ ratio to form negative training and testing samples, respectively. We computed inner product for all the edges and labeled them $0$ or $1$ depending on whether it is a missing edge or an actual edge. Then we combined the positive and negative test samples to form the test data. 

\hl{We learned embedding of the graph using all the candidate algorithms on our training samples.} To measure the performance of the classifier on the embeddings, we calculated the inner product similarity among the embeddings of a pair of nodes from the test data; {  for {\tt QUINT}, we used the \bs algorithm to estimate similarities}. We trained a logistic regression on the final training data considering \texttt{AUC-ROC} as our evaluation metric.

{\textbf{Datasets:}}
\texttt{Gowalla}~\cite{snap}  is a location-based social networking website where users share their locations by checking-in. 
The dataset consists of  a total of $6,442,890$ check-ins of these users over the period of Feb. 2009 - Oct. 2010. \texttt{Enron Emails Network}~\cite{snap}  covers all the email communication within a dataset of around half million emails. Nodes of the network are email addresses and if an address $i$ sent at least one email to address $j$, the graph contains an undirected edge from $i$ to $j$. 
\texttt{BlogCatalog}~\cite{BlogCatalog}  is the social blog directory which manages the bloggers and their blogs. 
The dataset contains the friendship network crawled and group memberships. Here, nodes represent users, and edges represent a friendship relation between any two users. 
In the \texttt{Facebook}~\cite{snap} network, nodes represent
users, and edges represent friendship relations between  users.

We used two social network datasets -- \texttt{Flickr} and \texttt{YouTube}~\cite{dataset}.
\texttt{Flickr} dataset is  of photo-sharing social network, where labels
represent the self-identified interests of particular users and edges correspond to the messages between two users.  
\texttt{YouTube} dataset is a video-based social network of user interactions where 
labels indicate interest in a particular video genre. We use these datasets, and {\tt BlogCatalog}, for both node classification and link prediction experiments. 
Statistics of these datasets are summarised in
Table~\ref{tab:data_link_prediction}.

{\textbf{Empirical results and insights:}}
Table~\ref{tab:link_prediction} summarizes the comparative analysis (Table~\ref{tab:link_prediction_full_128} in Supplementary has the complete details). We observed significant speedup for {\tt QUINT} in compression  time, along with better accuracy as compared to the other candidate algorithms.
For instance, on the Gowalla dataset  (using 1000 dimensions for {\tt QUINT}), we obtained $46\times$ to $10^4\times$ speedup on the compression times \textit{w.r.t.}
the other candidate algorithms. 
Our embedding also saves on space as well (see Table~\ref{tab:bits_space}).

It is evident that AUC-ROC for {\tt QUINT} is comparable with the best among the baselines (a tie on {\tt Gowalla} and {\tt Facebook} and a close second on {\tt Flickr} and {\tt Enron}), and sometimes even better (on {\tt Youtube, BlogCatalog}), but always using a fraction of their times. {\tt VERSE} turns out to be worthy alternative for all the datasets but {\tt Youtube}, but it is a lot slower because it learns an embedding by running a single-layer neural network.


{  An attractive property of $\binsketch$ is the ability to recover different types of similarities from its sketches, e.g., inner product, cosine similarity, etc. {\tt QUINT} inherits them too. We conducted link prediction experiments using three other similarity measures --- cosine similarity, $\ell_1$ and $\ell_2$ norm, and observed a similar trend that {\tt QUINT} offered significant speedup in the compression time while offering comparable AUC-ROC scores with respect to the baselines (see Table~\ref{tab:diff_similarity_measure} in Supplementary). This shows that {\tt QUINT} embeddings may be applicable for multiple similarity measures.}


\begin{table}
\centering
\caption{Datasets used for node classification. The first three datasets are also used for link prediction.}\label{tab:data_node_classfication}
\scalebox{1}{
\begin{tabular}{ |c|c|c|c|c|c| } 
 \hline

   Dataset & \# Nodes & \# Edges & \# Classes & Max. degree \\ 
 \hline
 { Flickr}   & 80,513 &  5,899,882 &195&  5706\\
 { Youtube}   & 1,138,499 & 2,990,443 & 35 & 28,754\\ 
 { BlogCatalog} & 10,312 & 333,983 & 39 & 3,992 \\ 
 IMDB  & 19,773 & 3,86,124 & 1000 & 540 \\
 Pubmed   & 19,717 & 44,338 & 3& 171 \\ 
  Citeseer    & 3,327 & 4,732 & 6& 26 \\ 
  Cora    & 2,708 & 5,429 & 7& 5 \\ 
 \hline
\end{tabular}}
\vspace{-5mm}
\end{table}

\begin{table*}
\centering
\caption{Comparison on Micro $F_1$, Macro $F_1$ and compressions time of \texttt{QUINT} and other baselines (using 128 dimensions) for the task of node classification. For the datasets on which {\tt QUINT} outperformed the baselines, we have reported the  performance of {\tt QUINT} on the smallest dimension at which it outperformed the latter; for the other datasets we have reported the best performance of {\tt QUINT} (see Tables~\ref{tab:node_classification1_full} and \ref{tab:blogcatalog_node_class} in Supplementary for the results for all the dimensions). 
We stopped baselines that took 10 hours or more and indicate them by \DNS; {\tt OOM} indicates baselines that ran out of memory. 
The best Macro $F_1$ scores among the baselines and for {\tt QUINT} are {\bf bolded}.
}
  \label{tab:node_classification1}
   \noindent \begin{tabular}{|l|cccc| cccc | cccc |}
    \toprule
    \multirow{1}{*}{Method} & {} &
      \multicolumn{3}{c|}{Flickr} &
      \multicolumn{4}{c|}{Youtube} & 
      \multicolumn{4}{c|}{IMDB}\\
      & {Dim.} & {Micro$F_1$} & {Macro$F_1$}&{Comp.}  & {Dim.} & {Micro$F_1$} & {Macro$F_1$}&{Comp.} & {Dim.} & {Micro$F_1$} & {Macro$F_1$}&{Comp.}\\ 
      &            & {Score}            &{Score}            &{ time(s)}   &            & {Score}            &{Score}            &{ time(s)}    &            & {Score}            &{Score}            &{ time(s)}\\
            \midrule
            \texttt{node2vec}& 128 & \NA[37.58] & \NA[21.05] & \DNS[44408.30 / 12.33 hr] 
                             & 128 & \NA[47.724] & \NA[41.214] & \DNS[39133.25 /10.87 hr]
                             & 128 & 51.91 & 51.6  & 117\\
            \texttt{deepwalk}& 128  & 41.2 & 29.61 & 2.1 hr 
                             & 128  & \NA[46.82] & \NA[39.37] & \DNS[100579.85 / 27.93 hr]
                             & 128 & 50.68 & 50.34 & 118\\
            \texttt{LINE} &128  & \NA[28.19] & \NA[13.96] & \DNS[266814.56 / 74 hr]
                              & 128 & \NA[29.47] & \NA[17.75] &  \DNS[1966862 / 546 hr]
                              & 128 & 41.25 & 40.56 & 1850\\
            \texttt{VERSE} &128  & 41.52 & {\bf 31.41} & 4.8 hr 
                           & 128 & \NA[47.80] & \NA[41.44] & \DNS[253484.53 / 70 hr]
                           & 128 & 63.09 & 63.08 & 1194.10\\
            \texttt{NetMF} &128 & \NA & \NA & \OOM 
                           & 128 & \NA & \NA & \OOM
                           & 128 & 60.30 & 51.88 & 21.72\\
            \texttt{NodeSketch} &128 & 22.37 & 7.1  & 1057.30 
                                &128 & 37.55 & {\bf 26.98} & 902.68 
                                & 128 & 55.69 & 46.64 & 6.98\\
            \hline 
            \texttt{SGH}& 128 & \NA & \NA & \OOM                                        & 128 & \NA & \NA & \OOM
            & 128 & 70.83 & 70.75 & 110.78\\
                        & 3000 & \NA & \NA & \OOM                                        & 128 & \NA & \NA & \OOM
                        & 3000 & 74.51 & {\bf 74.49} & 174.12\\
             \hline    
             {\tt QUINT} & 20000 & 37.04 & {\bf 29.99} & 59.89 & 3000 & 37.80 & {\bf 33.40} & 67.53 & 2000 & 83.88 & {\bf 83.85} & 3.89 \\
            \hline   
            \texttt{Uncompressed} &\NA  & \texttt{OOM} &   \texttt{OOM} & \NA     
            &\NA & \texttt{OOM} & \texttt{OOM} &  \NA
            & \NA  & 94.90 &   95.91 &  \hspace*{0.25cm} \NA       \hspace*{0.25cm}\\

     \bottomrule
  \end{tabular}\\[1em]
\addtolength{\tabcolsep}{-4pt}
    \noindent\begin{tabular}{|l|cccc|cccc| cccc| cccc|}
    \toprule
    \multirow{1}{*}{Method} &
      \multicolumn{4}{c|}{Citeseer} &
      \multicolumn{4}{c|}{Cora} & 
      \multicolumn{4}{c|}{Pubmed} &
      \multicolumn{4}{c|}{BlogCatalog} \\
      & {Dim.} & {Micro$F_1$} & {Macro$F_1$}&{Comp.}& {Dim.} & {Micro$F_1$} & {Macro$F_1$}&{Comp.} & {Dim.} & {Micro$F_1$} & {Macro$F_1$}&{Comp.} & {Dim.} & {Mic.$F_1$} & {Mac.$F_1$}&{Comp.}  \\ 
      &            & {Score}            &{Score}            &{ time(s)} &            & {Score}            &{Score}            &{ time(s)}&            & {Score}            &{Score}            &{ time(s)}&            & {Score}            &{Score}            &{ time(s)}\\
            \midrule
            \texttt{node2vec} & 128 & 29.77 & 21.90 & 1.75 
                              & 128 & 52.76 & 41.31 & 1.72
                              & 128 & 43.2  & 31.34 & 5.88
                              & 128 & 38.90 &  23.19 & 1359.25 \\
            \texttt{deepwalk}& 128  & 25.05 & 17.06 & 1.77
                             & 128  & 44.15 & 28.64 & 1.49
                             & 128 & 42.46 & 35.15 & 5.78
                             & 128 & 41.82  & 28.24 & 1167.95 \\
            \texttt{LINE} & 128 & 36.72 & 32.72 & 17.14
                              & 128 & 56.33 & 53.51 & 15.72
                              & 128 & 58.19 & 53.81 & 840.25
                              & 128 & 20.17 & 10.15 & 1898.23\\
            \texttt{VERSE} & 128 & 32.93 & 29.18 & 71.83 
                           & 128 & 52.27 & 46.70 & 53.2
                           & 128 & 42.29 & 31.93 & 806.98 
                           & 128 & 41.78 & 29.04  & 847.4 \\
            \texttt{NetMF} & 128 & 49.69 & \textbf{44.06} & 1.39
                           & 128 & 57.68 & 52.65 & 1.43
                           & 128 & 66.00 & {\bf 53.85} & 19.88
                            & 128 & 43.45 & {\bf 29.05} &  72.39 \\
            \texttt{NodeSketch} & 128 & 25.65 & 21.26 & 0.41
                                & 128 & 39.72 & 30.39 & 0.5
                                & 128 & 43.79 & 33.77 & 4.31
                                & 128 & 19.76 & 8.66 & 66.36 \\
            \hline 
            \texttt{SGH} & 128 & 37.77 & 33.37 & 2.79
                         & 128 & 56.45 & 52.96 & 2.54
                         & 128 & 51.10 & 44.80 & 67.67
                         & 128 & 13.20   & 4.05 & 18.25  \\
                        
                         & 2000 & 38.77 & 35.60 & 53.68
                         & 2000 & 57.93 & {\bf 56.07} & 36.33
                         & 2000 & 54.83 & 49.68 & 122.77
                         & 3000 & 9.02 & 4.74 & 120.27\\
             \hline 
             {\tt QUINT} & 3500 & 50.68 & {\bf 45.99} & 0.286 & 1000 & 61.62 & {\bf 60.43} & 0.055 & 2000 & 68.27 & {\bf 65.50} & 2.50 & 10000 & 37.70 & {\bf 25.51} & 2.35 \\
                            
             
                            
                           
                           
                            
                            
            \hline   
            \texttt{Uncompressed} &\NA & 51.40 & 47.06  & \hspace*{0.25cm} \NA \hspace*{0.25cm}
                                  & \NA & 65.75 & 66.66 & \hspace*{0.25cm} \NA \hspace*{0.25cm}
                                  &\NA & 76.52 & 74.98 &  \hspace*{0.25cm} \NA \hspace*{0.25cm}\
                                  &\NA  & 38.23 &   25.65 &  \hspace*{0.25cm} \NA   \hspace*{0.25cm}\\

     \bottomrule
  \end{tabular}\\ [1em]
\addtolength{\tabcolsep}{4pt}

\end{table*}

\subsection{Comparing performance on node classification}

For node classification, every node is assigned one or more labels from a given set. The dataset is randomly divided into 70\% and 30\% for training and testing, respectively. We repeated the experiment 10 times for each network and report the average accuracy. We used logistic regression as classifier. Unlike the experiments on link prediction where a subgraph with 70\% edges is embedded to generate training data, embedding of the {\em entire network} is first computed for the node classification experiments; this shows up in the form of larger compression times.

{\textbf{Datasets:}} { Some of the datasets used for link prediction are not labelled and hence, they are not suitable for node classification.} So we used three additional citation datasets
%
%
 --- \texttt{Cora}, {\tt Citeseer} and \texttt{Pubmed}~\cite{Sen08collectiveclassification} for the experiments. 
Here, citation relationships are viewed as directed  edges. Attributes associated with nodes are extracted from the title and the abstract of the each article and are presented as sparse bag-of-word vectors, after removing the stop words and low-frequency words. Each article in these datasets has only one label representing the class it belongs to. 
We also considered \texttt{IMDB-BINARY}~\cite{YanardagV15,imdb}, 
 which is a movie collaboration dataset where actor/actress and genre information of different movies on IMDB are collected. Nodes represent actors/actresses, and an edge between them signifies a joint appearance in some movie. Collaboration graphs is generated on the ``action'' and the ``romance'' \textit{genres} and \textit{ego-networks} are derived for each actor/actress. 
A movie can belong to both genres at the same time, therefore movies from the romance genre are discarded if they are already included in the action genre. 
Each ego-network is labeled with the genre graph it belongs to. The task is then simply to identify which genre an ego-network graph belongs to.
All the datasets are summarised in Table~\ref{tab:data_node_classfication}.

{\textbf{Empirical results and insights:}} 
Table~\ref{tab:node_classification1}  summarizes the performance of the competing algorithms (see Table~\ref{tab:node_classification1_full} for complete details) which show a similar trend as in the link prediction experiments. Many baselines failed to generate an embedding of {\tt Flickr} and/or {\tt Youtube} datasets within 10 hours, or ran out of memory; however, {\tt QUINT} did not have any such difficulty and was able to comfortably finish all embedding tasks within a few minutes, mostly taking a few seconds.

Compared with the baselines that could finish embedding within 10 hours, {\tt QUINT} achieved a higher Macro F1 on {\tt Youtube, IMDB, Citeseer, Cora, Pubmed} and was a close second on {\tt Flickr}. It fell behind by 10-15\% on {\tt BlogCatalog} which could be due its network characteristics; even {\tt Uncompressed}, that which {\tt QUINT} tries to improve upon, was unable to obtain a high score on that network. The general conclusion that we can draw is that {\tt QUINT} offers significant speed-up during node embedding while offering comparable accuracy for node classification when compared with the state-of-the-art techniques.


\subsection{Comparing \texttt{Uncompressed} with \texttt{QUINT.}}
{ We also want to draw attention towards the {\tt Uncompressed} embedding. However na\"ive it may sound, empirically it appears that the neighborhood of a node, when represented as an array, is a reasonably good embedding for the purposes of link prediction and node classification (see the last rows of Tables~\ref{tab:link_prediction} and \ref{tab:node_classification1}). The challenge is that training on such embeddings of a large network (e.g., {\tt Flickr} with 80K nodes) is almost impossible on off-the-shelf hardware. Even for smaller networks, the training time is significantly higher compared to the {\tt QUINT} embeddings. We list out the speed-up in training time of \texttt{QUINT} over \texttt{Uncompressed} in Table~\ref{tab:qunitvsadjacency}. {\tt QUINT} embeddings are a compressed version of {\tt Uncompressed} embeddings, and we can safely conclude that it is able to strike a good balance between speed and accuracy.}



\begin{table*}
\caption{ Comparison on speedup in training time on embeddings generated using \texttt{QUINT} {\it vs.} \texttt{Uncompressed}. We choose the embedding dimension of \texttt{QUNIT} on which it outperforms (or reaches sufficiently close to) the best baselines.}
\vspace{-5mm}
\begin{center}
\begin{tabular}{ |c|c|c|c|c|c|c|c|c|c|c|c| } 
 \hline
  Dataset&Gowalla& Flickr & Youtube &Enron & \multicolumn{2}{c|}{BlogCatalog}& Facebook &IMDB &Pubmed &Citeseer& Cora \\
  & & & & & Link pred. & node class.& & & & & \\
  \hline
  Dim. of \texttt{QUINT} &1000& 100 & 100 & 2000 & 100 & 5000 & 1000 & 1000& 1000& 2000 & 1000\\
  Speedup (training time) &\texttt{OOM} &\texttt{OOM} & \texttt{OOM}& 96.16$\times$ & 86.4$\times$ & 24.8$\times$ & 31.25$\times$ & 77.65$\times$ & 10.48$\times$ & 2.80$\times$  & 4.92$\times$ \\
 \hline
\end{tabular}
 \end{center}
\label{tab:qunitvsadjacency}
\vspace{-5mm}
\end{table*}


\subsection{Performance on training size}\label{subsection:var_sparsity}
\begin{figure}[ht!]
\centering
\includegraphics[width=0.5\textwidth]{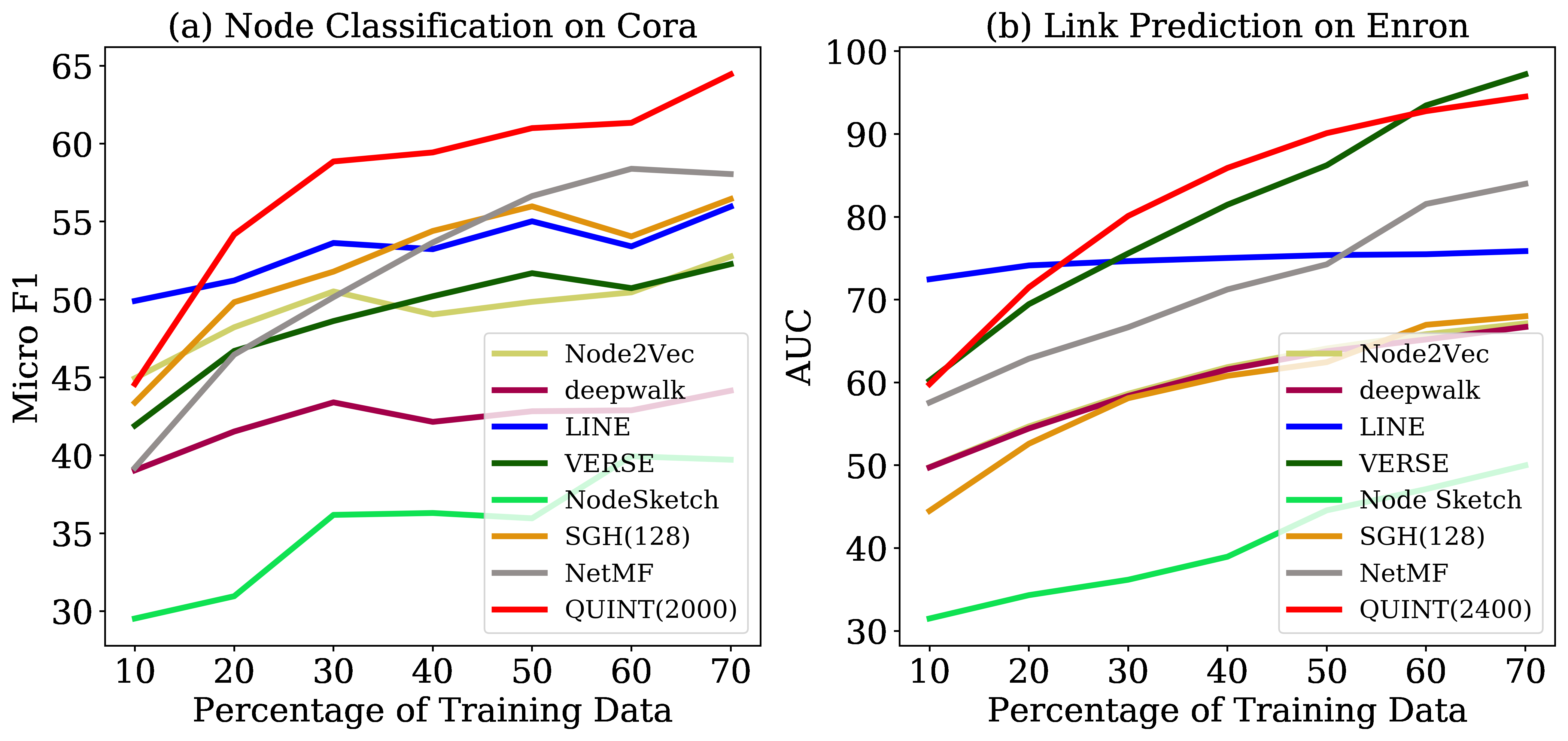}
\vspace{-2mm}
\caption{Performance of the baselines on various  training/test splits. Figure (a) summarises the Micro $F_1$ scores \textit{vs} $\%$ of training data  of various  baselines on node classification for the Cora dataset. Figure (b) summarises the AUC-ROC scores \textit{vs} $\%$ of training data of various baselines on link prediction for the Enron dataset. }
\label{fig:multiple_split_node_classification}
\end{figure}

We varied the training size in our experiments to show how the competing algorithms react to this variation. For this we split the datasets into various ratios of training and test partitions, starting at $10\%$ training and $90\%$ test partitions, and moving to $90\%$ training and $10\%$ test partitions at intervals of $10\%$. We ran node classification and link prediction on these partitions and observed the Micro F1 and AUC-ROC scores, respectively. We compared the performance of \texttt{QUINT} \textit{w.r.t.} other candidate algorithms for these experiments, and report the results for two datasets in Figure~\ref{fig:multiple_split_node_classification}. The results for the other datasets were similar.

We noticed that for node classification on the \textit{Cora} dataset the Micro $F_1$ score always remains higher as compared to the other candidate algorithms (we obtained a similar result on Macro $F_1$ score as well). For link prediction on the \textit{Enron} dataset, we noticed that the AUC-ROC score is comparable to that of {\tt VERSE} and significantly better than the other baselines. This essentially confirms that even on small training data, \texttt{QUINT} maintains its competitive advantage \textit{w.r.t.} other baselines.


\subsection{Scalability experiments}
\begin{figure}[t]
\centering
\includegraphics[width=0.5\textwidth]{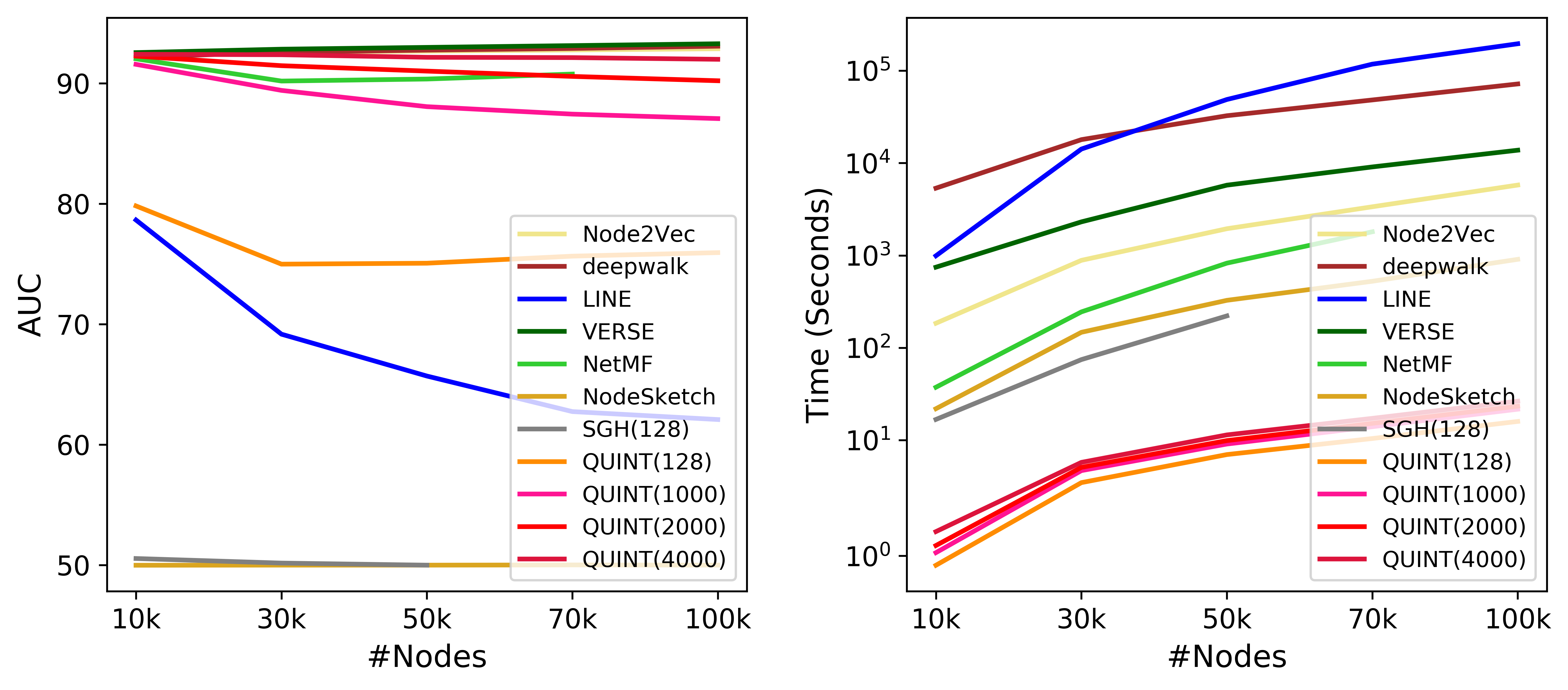}
\vspace{-2mm}
\caption{Comparing link prediction performance of {\tt QUINT} and the baseline algorithms on LFR graphs by varying the number of nodes in a network. The left plot compares the AUC-ROC scores and the right plot compares the corresponding compression times. SGH gives \texttt{OOM} error for $70K$ and $100K$ nodes, and NetMF gives \texttt{OOM} error for $100K$  nodes. }
\label{fig:varying_nodes}
\vspace{-5mm}
\end{figure}

\begin{figure}[b]
\centering
\includegraphics[width=0.5\textwidth]{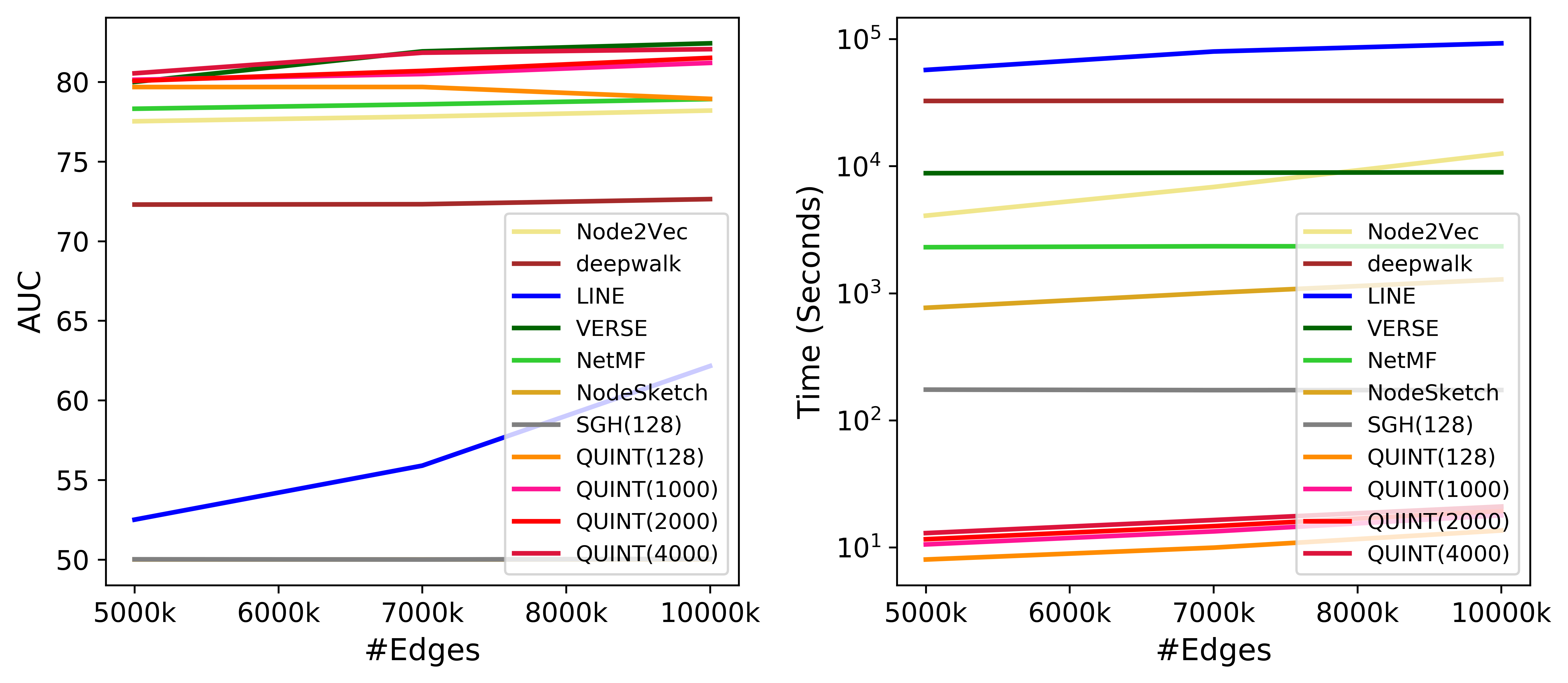}
\vspace{-2mm}
\caption{Comparing link prediction performance of {\tt QUINT} and the baseline algorithms on LFR graphs by varying the number of edges in a network with 50K nodes. The left plot compares the AUC-ROC scores and the right plot compares the corresponding compression times. }
\label{fig:varying_edges}
\vspace{-5mm}
\end{figure}

{ 
Due to the bitwise operations and simplicity of Algorithm~\ref{algo:QUINT}, the time taken by {\tt QUINT} increases very slowly in the embedding dimension. This was evident in our experiments on the real-life datasets (see  Section~\ref{sec:dimension_varying} and  Tables~\ref{tab:node_classification1_full} and \ref{tab:link_prediction_full_128} in Supplementary).

To evaluate the effect of the size of a network, we experimented with link prediction on synthetically generated LFR graphs~\cite{Lancichinetti2008Benchmark} of varied nodes and edges generated using Python NetworkX with parameters $\mu=0.1,\tau_1=2, \tau_2=1.1$. 
The first set of experiments were on graphs with 10K to 100K nodes and second were on graphs with 50K nodes and 1000K to 10,000K edges; the AUC-ROC scores and compression times of {\tt QUINT} and the baseline algorithms are plotted in Figures~\ref{fig:varying_nodes} and \ref{fig:varying_edges}, respectively. We observed that in both the experiments, {\tt QUINT}, when embedded on $1000 ~-~4000$ dimensions, is consistently producing a high AUC-ROC along with {\tt node2vec, deepwalk, VERSE} and {\tt netmf}. Few of the baselines, including {\tt netmf}, did not run for the graph with 1000K nodes, and {\tt QUINT} exhibited one to four orders of speedup for the rest.
}

 \section{Conclusion and open questions}\label{sec:Conclusion}
In this work, we proposed \texttt{QUINT} which takes a large scale graph as input and outputs succinct low-dimensional binary embedding for each node. The major advantage of \texttt{QUINT} is that it is extremely fast -- it computes the embedding of a large  graph in almost real time. \texttt{QUINT} does not have a strong hardware requirement and consumes less space for computing and storing the embedding of a graph. In fact, most of our experiments were conducted on an off-the-shelf laptop.
We evaluated the performance of \texttt{QUINT} on the task of \textit{node classification} and \textit{link prediction} and noticed that  \texttt{QUINT} offers massive speed up in compression time while offering comparable performance with respect to the state-of-the-art algorithms. Moreover, our embedding is binary which makes it space efficient as compared to the real-valued embeddings generated by the most of the baselines.


\quint has a few added advantages that could be explored further. First is applicability in the sense that the same embedding can be used for both link prediction and node classification, and we conjecture that the same would be applicable to other machine learning tasks based on structural properties of networks, like node clustering. The second advantage comes from the bitwise nature of \quint. It maybe possible to use \quint in a distributed setting where embeddings of different groups of nodes are generated on different machines and then combined. Observe that the combination of multiple \quint embeddings is simply their bitwise-{\tt OR}.

The final advantage is in regards to evolving networks. We do not consider them in this work but we do make the observation that \quint embeddings are easy to update. This is since an addition of a new edge, say between nodes $i$ and $j$, essentially leads to setting one bit each in the embeddings of $i$ and $j$ and this can be done independently. To the best of our understanding, most learning-based algorithms would require solving an optimization problem afresh which may mean running their entire algorithm once again.

Our work leaves open the possibility of several future directions. First, we wonder if our results can be extended to efficiently embed hypergraphs? Hypergraphs are a common choice to model non-binary relationships but are difficult to analyse due to high complexity of even simple tasks. A few solutions have been proposed for clustering, classification, and other data analytic tasks based on spectral techniques~\cite{zhou2007learning} and random walk~\cite{yang2020lbsn2vec_tkde}. A binary embedding technique {\it ala.} \quint would make hypergraphs considerably easy to handle.

On the theoretical side it would be worth exploring what specific structural features of networks can be efficiently estimated from their embeddings. For example, global clustering coefficient of nodes could be one such candidate since it is basically the ratio of the number of triangle and an expression involving degree of each node. The number of triangles involving a pair of nodes, say $u$ and $v$, can be written as $n_{u,v} \cdot E_{u,v}$, in which $n_{u,v}$ denotes the number of other nodes that are connected to both $u$ and $v$, and $E_{u,v}$ is an indicator variable denoting the presence of an edge between $u$ and $v$. We have shown earlier how to estimate $n_{u,v}$ (Theorem~\ref{thm:binsketch}) and $E_{u,v}$ (Lemma~\ref{lem:first-order}). Thus we conjecture that it should be possible to estimate global clustering coefficient with reasonable accuracy.

There is obviously the question of what other practical applications \quint, rather $\binsketch$, is capable of? One way to extend \quint would be by making it use the attributes associated with nodes and edges of a graph. Randomization may once again prove beneficial but we do not have a solution in sight. Given the affinity of \quint towards sparse graphs, we wonder whether \quint can be successful for graphs that are sparse for individual features but dense otherwise. We hope that our simple yet powerful technique can be easily adapted to bring forth efficient solutions to challenging graph embedding problems.

\ifCLASSOPTIONcompsoc
  \section*{Acknowledgments}
\else
  \section*{Acknowledgment}
\fi
The authors would like to thank Raghav Kulkarni for some discussions during the initial phase of work and Karthik Revanuru for some initial experiments. T.~Chakraborty would like to thank Ramanujan Fellowship (SERB),  DST (ECR/2017/00l691) and ihub-Anubhuti-iiitd Foundation set up under the NM-ICPS scheme of the Department of Science
and Technology, India.

{\small

}


\vspace*{5mm}

\begin{IEEEbiography}[{\includegraphics[width=1in,height=1.25in,clip,keepaspectratio]{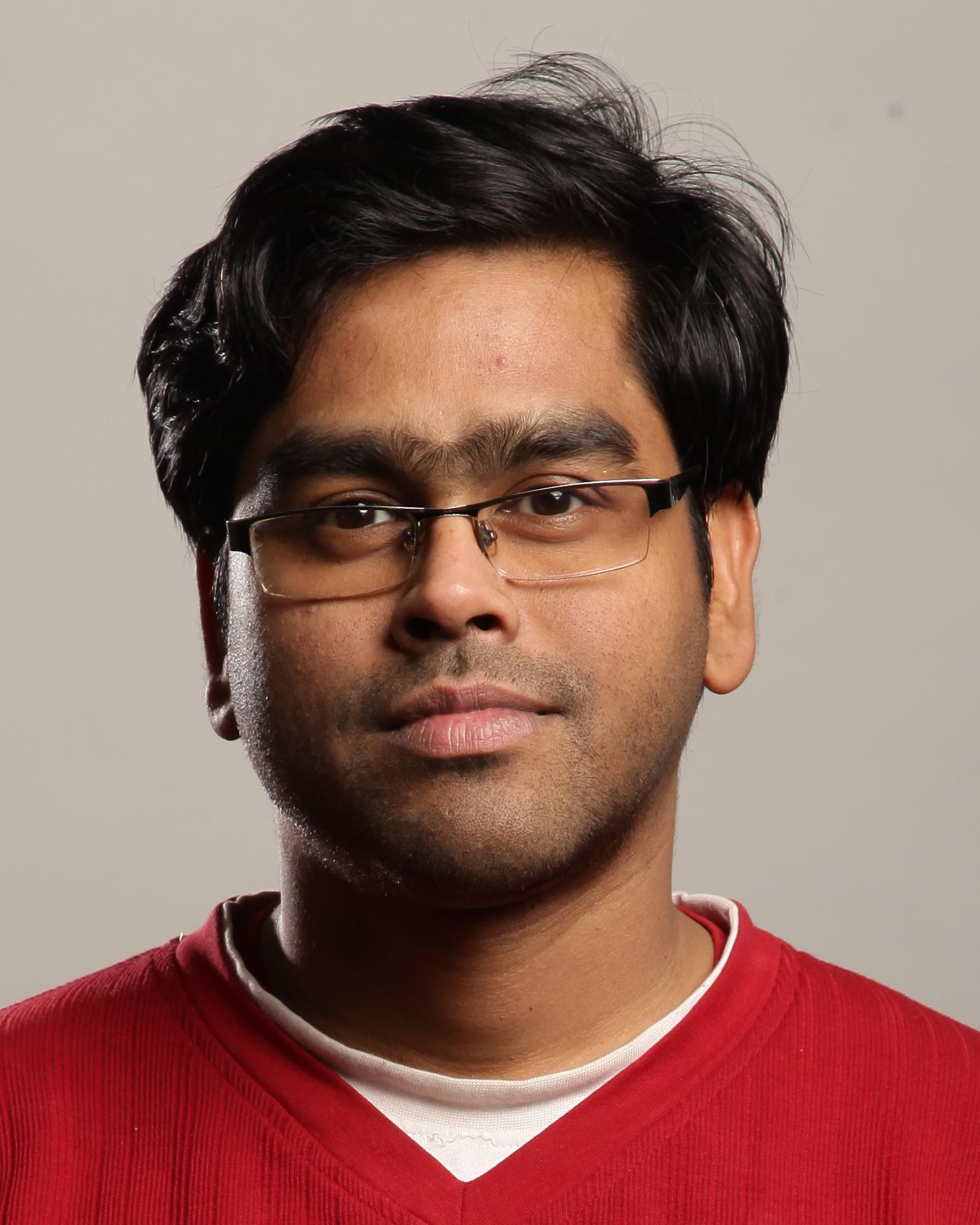}}]{Debajyoti~Bera}
received his B.Tech. in Computer Science and Engineering in 2002 at Indian Institute of Technology (IIT), Kanpur, India and his Ph.D. degree in Computer Science from Boston University, USA in 2010. Since 2010 he is an assistant professor at Indraprastha Institute of Information Technology, (IIIT-Delhi), India.
His research interests include quantum computing, randomized algorithms, and engineering algorithms for networks, data mining, and information security.
\end{IEEEbiography}
\vspace{-4mm}

\begin{IEEEbiography}[{\includegraphics[width=1in,height=1.25in,clip,keepaspectratio]{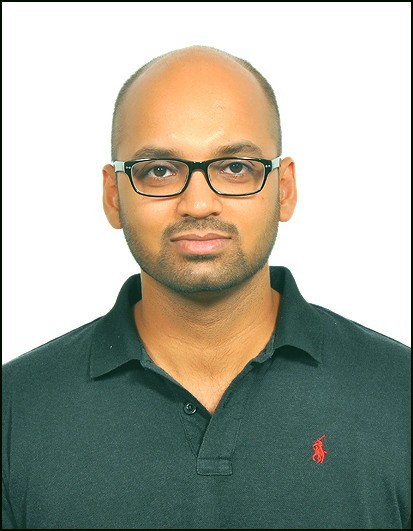}}]{Rameshwar Pratap}
 has earned Ph.D in Theoretical Computer Science in 2014  from Chennai Mathematical Institute (CMI). Earlier, he completed Masters in Computer Application (MCA) from  Jawaharlal Nehru University and   BSc in Mathematics, Physics, and Computer Science from University of Allahabad. Post Ph.D he has worked  TCS Innovation Labs (New Delhi, India),  and Wipro AI-Research (Bangalore, India). Since 2019 he is working as an assistant professor at School of Computing and Electrical Engineering (SCEE), IIT Mandi. His research interests include algorithms for dimensionality reduction,  robust sampling, and algorithmic fairness. 
\end{IEEEbiography}

\vspace{-4mm}

\begin{IEEEbiography}[{\includegraphics[width=1in,height=1.25in,clip,keepaspectratio]{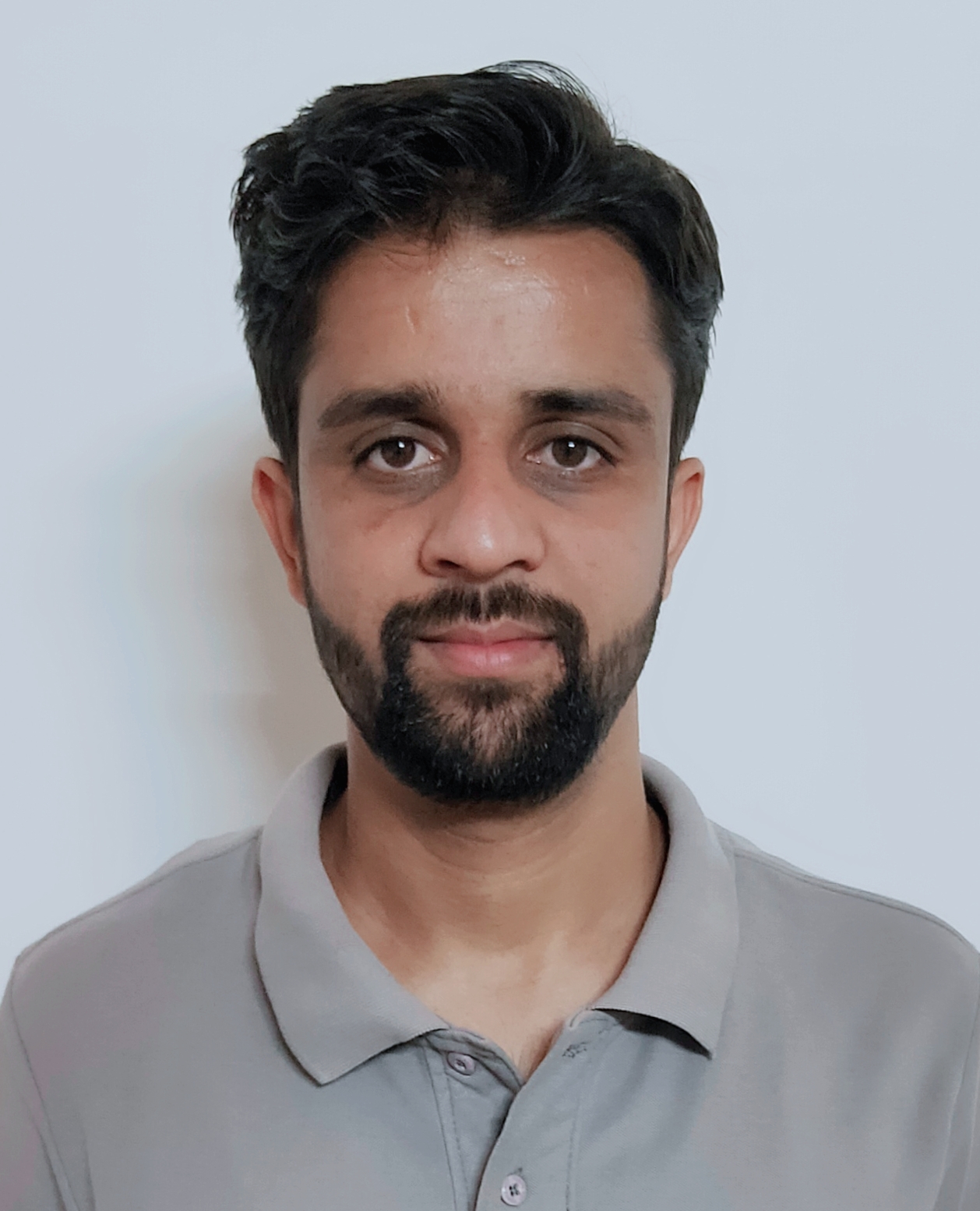}}]{Bhisham Dev Verma}
 is pursuing Ph.D from IIT Mandi. He has done his Masters in Applied Mathematics from IIT Mandi and BSc in Mathematics, Physics, and Chemistry from Himachal Pradesh University. His research interest includes data mining, algorithms for dimension reduction, optimization and machine learning.
 \end{IEEEbiography}
 
 \vspace{-4mm}

\begin{IEEEbiography}[{\includegraphics[width=1in,height=1.25in,clip,keepaspectratio]{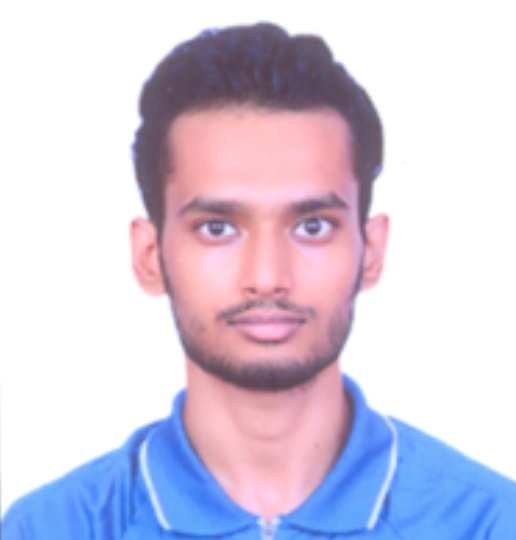}}]{Biswadeep Sen}
 is currently working as a research associate (remotely) at department of  Computer Science, National University of Singapore (NUS), since July, 2020. He completed  M.Sc. in
Computer Science and   B.Sc.  in Mathematics and Computer Science from Chennai  Mathematical  Institute  (CMI)  in  2018 and 2016, respectively. His research interests include Machine Learning, Artificial Intelligence, Computer Vision and Computational Sustainability.
\end{IEEEbiography}
 \vspace{-4mm}


\begin{IEEEbiography}[{\includegraphics[width=1in,height=1.25in,clip,keepaspectratio]{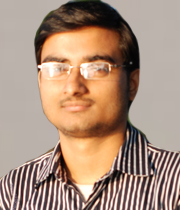}}]{Tanmoy Chakraborty}
is an Assistant Professor and a Ramanujan Fellow at the Dept. of CSE, IIIT-Delhi, India, where he leads \href{http://lcs2.iiitd.edu.in/}{Laboratory for Computational Social Systems (LCS2)}. His primary research interests include Social Network Analysis, Data Mining, and Natural Language Processing. He has received several awards including  Faculty Awards from Google, IBM and Accenture; Early Career Research Award, DAAD Faculty Fellowship.
\end{IEEEbiography}

\newpage

\onecolumn

\twocolumn[{\centering\textbf{\huge \texttt{QUINT}: Node embedding using network hashing (Appendix)}}]

\appendices

  \begin{figure}[!h]
\centering
\includegraphics[width=\columnwidth]{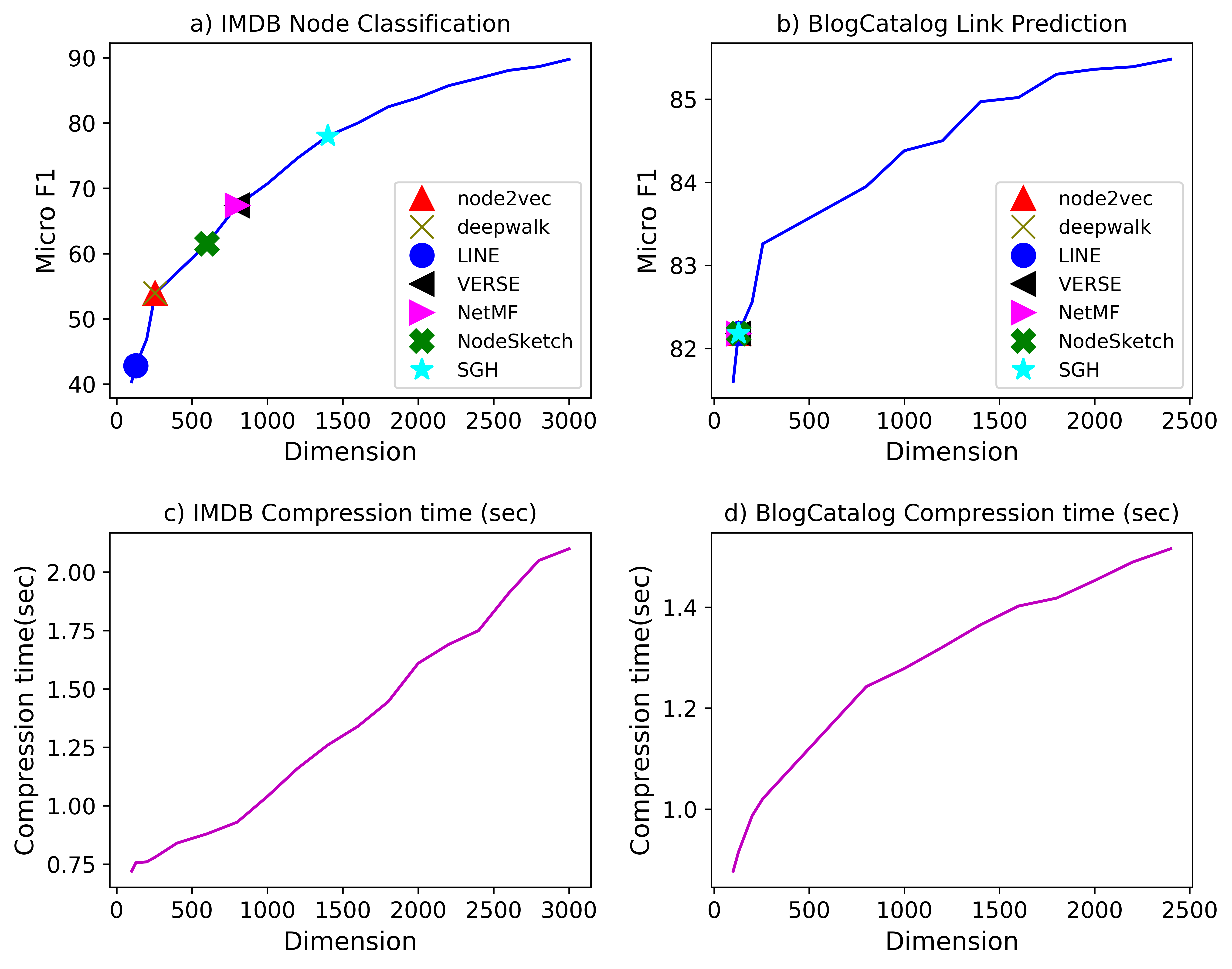}
\caption{{ 
Figure (a) shows the dimensions at which {\tt QUINT}'s node classification performance on the IMDB dataset matches that of various baselines, and Figure (c) shows the time {\tt QUINT} takes to do so. Figures (b) and (d) show the same for link prediction on the BlogCatalog dataset.}}
\label{fig:quint_multiple_dim}
\end{figure}

\section{Effect of embedding dimension}\label{sec:dimension_varying}
{ The optimum embedding dimension for {\tt QUINT} on any dataset is theoretically upper-bounded by an expression on its sparsity (the largest degree of a node). However, it is evident from Tables~\ref{tab:link_prediction} and \ref{tab:node_classification1} that much lower dimensions works in practice. Further, we observed that embedding on a higher dimension is not much slower and requires at most a few additional seconds (see Tables~\ref{tab:link_prediction_full_128} and \ref{tab:node_classification1_full} in Supplementary for the compression times on all dimensions).

To illustrate this observation further, we plot the performance and compression times of {\tt QUINT} against embedding dimension for the {\tt IMDB} dataset and the {\tt BlogCatalog} dataset, for node classification and link prediction, respectively. Figures \ref{fig:quint_multiple_dim}(a) and \ref{fig:quint_multiple_dim}(c) show the embedding dimensions of QUINT at which it matches the performance of the other baselines.  Figures \ref{fig:quint_multiple_dim}(b) and \ref{fig:quint_multiple_dim}(d) show the times taken by QUINT when it matches the baselines. These figures show that QUINT is able to match the performance of the baselines for node classification on IMDB and link prediction on BlogCatalog, and that too in only a few seconds (similar figures can be inferred from Tables~\ref{tab:link_prediction} and \ref{tab:node_classification1} for the other datasets and baselines that QUINT matches).}



\begin{table*}
  
\centering
\caption{Result of link prediction on the BlogCatalog dataset where instead of inner product, we use few other  similarity measures.
}
  \begin{tabular}{|l|ccc| ccc| ccc|}
    \toprule
    \multirow{1}{*}{Method} &
      \multicolumn{3}{c|}{BlogCatalog (cosine similarity)} &
      \multicolumn{3}{c|}{BlogCatalog ($\ell_1$ norm)} &
      \multicolumn{3}{c|}{BlogCatalog ($\ell_2$ norm)} \\
         & {Dim.} & {AUC-ROC} &{Time (s)}  & {Dim.} & {AUC-ROC} &{Time (s)}  & {Dim.} & {AUC-ROC} &{Time (s)} \\

            \midrule
     \texttt{node2vec} 
                       & 128  & 65.93 & 979.02   
                       & 128  & 75.09 & 979.02 
                       & 128  & 75.36 & 979.02  \\
        \texttt{deepwalk} 
                          & 128  & 61.6   & 774.31
                          & 128  & {\bf 82.66}  & 774.31
                          & 128  & {\bf 82.83}  & 774.31 \\
                          
        \texttt{LINE}   
                           & 128 & 65.07 & 2379
                           & 128 & 65.15 & 2379
                           & 128 & 36.79 & 2379 \\
                           
       \texttt{VERSE} 
                      & 128 & {\bf 79.3}  & 351.38
                      & 128 & 61.32 & 351.38
                      & 128 & 62.47 & 351.38\\
         
         \texttt{NetMF}   
                        & 128 & 62.18 & 7.7  
                        & 128 & 79.54 &  7.7
                        & 128 & 79.84 & 7.7\\
                        
         \texttt{NodeSketch} 
                             & 128  & 61.12 & 75.53 
                             & 128  & 50.00   & 75.53
                             & 128  & 50.00   & 75.53\\
         \hline
         
          \texttt{SGH} 
                       & 128 &   49.95   & 7.53
                       & 128 &   49.99   & 7.53
                       & 128 &   49.95   & 7.53\\

         \hline
         \texttt{QUINT} 
                        & 100 & {\bf 84.88} & 0.88
                        & 100 & 65.65 & 0.88
                        & 100 & 65.74 & 0.88\\

                        & 128 & 84.77& 0.92
                        & 128 & 72.18 & 0.92
                        & 128 & 72.82 & 0.92 \\

                        & 200 & 84.10 & 0.99
                        & 200 & 80.98 & 0.99
                        & 200 & 81.16 & 0.99\\

                        & 256 & 83.87 & 1.02
                        & 256 & 83.8  & 1.02
                        & 256 & 84.24 & 1.02\\

                        & 1000 & 79.69 & 1.28
                        & 1000 & {\bf 85.48} &  1.28
                        & 1000 & 86.4  & 1.28 \\

                        & 2000 & 77.3 & 1.45
                        & 2000 & 85.22 & 1.45
                        & 2000 & {\bf 86.36} & 1.45 \\

                        & 3000 & 76.18 & 1.62
                        & 3000 & 85.07 & 1.62
                        & 3000 & 86.31 &  1.62 \\

                        & 4000 & 75.29 & 1.77
                        & 4000 & 85.02 & 1.77
                        & 4000 & 86.29 & 1.77 \\
                         
         \hline
         \texttt{Uncompressed}     
                                 & \NA & 74.63 & \NA  
                                 & \NA & 84.60 & \NA 
                                 & \NA & 86.16 & \NA\\
    \bottomrule
  \end{tabular}
  \label{tab:diff_similarity_measure}
  \end{table*}

\begin{table*}
  
\centering
\caption{Comparison on Micro $F_1$, Macro $F_1$ and compressions time of \texttt{QUINT} and other baselines (using 128 dimensions) for the task of node classification.
We stopped baselines that took 10 hours or more and indicate them by \DNS; {\tt OOM} indicates baselines that ran out of memory. Smaller datasets were embedded to a maximum of 3000 dimensions. 
The best Macro $F_1$ scores among the baselines and for {\tt QUINT} are {\bf bolded}.
}
  \label{tab:node_classification1_full}
   \noindent \begin{tabular}{|l|c|ccc| Hccc | cccc |}
    \toprule
    \multirow{1}{*}{Method} & {} &
      \multicolumn{3}{c|}{Flickr} &
      \multicolumn{4}{c|}{Youtube} & 
      \multicolumn{4}{c|}{IMDB}\\
      & {Dim.} & {Micro$F_1$} & {Macro$F_1$}&{Comp.}  & {Dim.} & {Micro$F_1$} & {Macro$F_1$}&{Comp.} & {Dim.} & {Micro$F_1$} & {Macro$F_1$}&{Comp.}\\ 
      &            & {Score}            &{Score}            &{ time(s)}   &            & {Score}            &{Score}            &{ time(s)}    &            & {Score}            &{Score}            &{ time(s)}\\
            \midrule
            \texttt{node2vec}& 128 & \NA[37.58] & \NA[21.05] & \DNS[44408.30 / 12.33 hr] 
                             & 128 & \NA[47.724] & \NA[41.214] & \DNS[39133.25 /10.87 hr]
                             & 128 & 51.91 & 51.6  & 117\\
            \texttt{deepwalk}& 128  & 41.2 & 29.61 & 2.1 hr 
                             & 128  & \NA[46.82] & \NA[39.37] & \DNS[100579.85 / 27.93 hr]
                             & 128 & 50.68 & 50.34 & 118\\
            \texttt{LINE} &128  & \NA[28.19] & \NA[13.96] & \DNS[266814.56 / 74 hr]
                              & 128 & \NA[29.47] & \NA[17.75] &  \DNS[1966862 / 546 hr]
                              & 128 & 41.25 & 40.56 & 1850\\
            \texttt{VERSE} &128  & 41.52 & {\bf 31.41} & 4.8 hr 
                           & 128 & \NA[47.80] & \NA[41.44] & \DNS[253484.53 / 70 hr]
                           & 128 & 63.09 & 63.08 & 1194.10\\
            \texttt{NetMF} &128 & \NA & \NA & \OOM 
                           & 128 & \NA & \NA & \OOM
                           & 128 & 60.30 & 51.88 & 21.72\\
            \texttt{NodeSketch} &128 & 22.37 & 7.1  & 1057.30 
                                &128 & 37.55 & {\bf 26.98} & 902.68 
                                & 128 & 55.69 & 46.64 & 6.98\\
            \hline 
            \texttt{SGH}& 128 & \NA & \NA & \OOM                                        & 128 & \NA & \NA & \OOM
            & 128 & 70.83 & 70.75 & 110.78\\
                        & 3000 & \NA & \NA & \OOM                                        & 128 & \NA & \NA & \OOM
                        & 3000 & 74.51 & {\bf 74.49} & 174.12\\
             \hline    
             \texttt{QUINT} & 100 & 19.59 & 5.4 & 20.86
                            & 100 & 28.66 & 16.65 & 17.90
                            & 100 & 40.37 & 39.78 & 0.72 \\
                            
                            & 128 & 21.08 & 7.3 & 23.11
                            & 128 & 30.38 & 20.10 & 20.46
                            & 128 & 42.82 & 41.35 & 0.756 \\
                            
                            & 200 & 23.58 & 10.68 & 23.88
                            & 200 & 32.31 & 22.76 & 20.80
                            & 200 & 46.91 & 46.48 & \textbf{0.76} \\
                            
                            & 256 & 25.41 & 12.46 & 25.80
                            & 256 & 32.67 & 24.10 & 26.74 
                            & 256 & 53.89 & 52.45 & 0.78 \\
                            
                            & 1000 & 28.04 & 18.61 & 30.63              
                            & 1000 & 35.22 & 31.63 & 35.06
                            & 1000 & 70.7 & 70.58 & 1.04 \\
                            
                            & 2000 & 30.12 & 21.76 & 33.65              
                            & 2000 & 36.56 & 31.69 & 51.57
                            & 2000 & 83.88 & 83.85 & 1.61 \\
                            
                            & 3000 & 31.52 & 23.58 & 36.07              
                            & 3000 & 37.80 & 33.40 & 67.53
                            & 3000 & 89.76 & \textbf{89.81} & 2.1 \\
                            
                            & 4000 & 32.35 & 25.05 & 38.47              
                            & 4000 & 38.93 & 35.12 & 83.37
                            &      &       &       &      \\
                            
                            & 6000 & 33.98 & 26.54 & 41.85           
                            & 6000 & 39.61 & 35.89 & 111.52 
                            &      &       &       &      \\
                            
                            & 8000 & 34.66 & 27.31 & 44.86         
                            & 8000 & 40.94 & 37.06 & 140.08
                            &      &       &       &      \\
                            
                            & 10000 & 35.08 & 28.10 &  46.87             
                            & 10000 & 41.20 & 37.46 & 167.94 
                            &      &       &       &      \\
                            
                            & 16000 & 36.86 & 29.75 & 55.23          
                            & 16000 & 42.41 & 38.01 & 250.03
                            &      &       &       &      \\
                            
                            & 20000 & 37.04 & {\bf 29.99} & 59.89              
                            & 20000 & 43.26 & {\bf 38.68} &  325.48
                            &      &       &       &      \\
            \hline   
            \texttt{Uncompressed} & \NA  & \texttt{OOM} &   \texttt{OOM} & \NA     
            & \NA & \texttt{OOM} & \texttt{OOM} &  \NA
            & \NA  & 94.90 &   95.91 &  \hspace*{0.25cm} \NA       \hspace*{0.25cm}\\

     \bottomrule
  \end{tabular}\\[1em]

    \noindent\begin{tabular}{|l|c|ccc|Hccc| Hccc|}
    \toprule
    \multirow{1}{*}{Method} &
      \multicolumn{4}{c|}{Citeseer} &
      \multicolumn{4}{c|}{Cora} & 
      \multicolumn{4}{c|}{Pubmed} \\
      & {Dim.} & {Micro$F_1$} & {Macro$F_1$}&{Comp.}& {Dim.} & {Micro$F_1$} & {Macro$F_1$}&{Comp.} & {Dim.} & {Micro$F_1$} & {Macro$F_1$}&{Comp.}  \\ 
      &            & {Score}            &{Score}            &{ time(s)} &            & {Score}            &{Score}            &{ time(s)}             & {Score}            &{Score}            &{ time(s)}\\
            \midrule
            \texttt{node2vec} & 128 & 29.77 & 21.90 & 1.75 
                              & 128 & 52.76 & 41.31 & 1.72
                              & 128 & 43.2  & 31.34 & 5.88\\
            \texttt{deepwalk}& 128  & 25.05 & 17.06 & 1.77
                             & 128  & 44.15 & 28.64 & 1.49
                             & 128 & 42.46 & 35.15 & 5.78\\
            \texttt{LINE} & 128 & 36.72 & 32.72 & 17.14
                              & 128 & 56.33 & 53.51 & 15.72
                              & 128 & 58.19 & 53.81 & 840.25\\
            \texttt{VERSE} & 128 & 32.93 & 29.18 & 71.83 
                           & 128 & 52.27 & 46.70 & 53.2
                           & 128 & 42.29 & 31.93 & 806.98 \\
            \texttt{NetMF} & 128 & 49.69 & \textbf{44.06} & 1.39
                           & 128 & 57.68 & 52.65 & 1.43
                           & 128 & 66.00 & {\bf 53.85} & 19.88\\
            \texttt{NodeSketch} & 128 & 25.65 & 21.26 & 0.41
                                & 128 & 39.72 & 30.39 & 0.5
                                & 128 & 43.79 & 33.77 & 4.31\\
            \hline 
            \texttt{SGH} & 128 & 37.77 & 33.37 & 2.79
                         & 128 & 56.45 & 52.96 & 2.54
                         & 128 & 51.10 & 44.80 & 67.67\\
                        
                         & 2000 & 38.77 & 35.60 & 53.68
                         & 2000 & 57.93 & {\bf 56.07} & 36.33
                         & 2000 & 54.83 & 49.68 & 122.77\\
             \hline    
             \texttt{QUINT} & 100 & 34.3 & 28.42 & 0.024
                            & 100 & 44.65 & 38.58 & 0.02
                            & 100 & 47.71 & 39.27 & 0.2\\
                            
                            & 128 & 34.95 & 28.75 & 0.029 
                            & 128 & 46.37 & 41.71 & 0.021
                            & 128 &  48.63& 42.55 &0.21 \\
             
                            & 200 & 36.51 & 28.98 &  0.035
                            & 200 & 51.78 & 47.87 & 0.024
                            & 200 & 50.62 & 43.12 & 0.23\\
                            
                            & 256 & 38.55 & 29.80 & 0.038
                            & 256 & 54.12 & 50.55 & 0.027 
                            & 256 & 52.96 & 47.58 & 0.25 \\
                           
                            & 1000 & 44.16 & 38.79 & 0.076 
                            & 1000 & 61.62 & 60.43 & 0.055
                            & 1000 & 63.28 & 59.63 & 1.23 \\
                           
                            & 2000 & 48.28 & 43.16  &  0.112 
                            & 2000 & 63.59 & 61.64 & 0.094
                            & 2000 & 68.27 & 65.59 & 2.50\\
                            
                            & 3000 & 49.15 & 43.98 & 0.214  
                            & 3000 &  64.945 & {\bf 64.32} & 0.124 
                            & 3000 & 69.25 & {\bf 66.32} & 3.75 \\
                            
                            & 3500 & 50.68 & {\bf 45.99} & 0.286
                            & & & & \\
            \hline   
            \texttt{Uncompressed} &\NA & 51.40 & 47.06  & \hspace*{0.25cm} \NA \hspace*{0.25cm}
                                  & \NA & 65.75 & 66.66 & \hspace*{0.25cm} \NA \hspace*{0.25cm}
                                   &\NA & 76.52 & 74.98 &  \hspace*{0.25cm} \NA \hspace*{0.25cm}\\

     \bottomrule
  \end{tabular}\\ [1em]
\end{table*}

\def \ab{11}
\def \bc{10}

\ifx\ab\bc{
\begin{table*}
  
\centering
\caption{Comparison on Micro and Macro $F_1$ scores  and  compression time of \texttt{QUINT} and other baselines on the task for node classification. \texttt{Uncompressed} denotes running node classification algorithm on the full adjacency matrix  of the graph. 
We mark the best two Micro/Macro $F_1$ scores  and the fastest compression time in boldface.}
  \label{tab:node_classification1}
   \noindent \begin{tabular}{|l|cccc| cccc |}
    \toprule
    \multirow{1}{*}{Method} &
      \multicolumn{4}{c|}{Flickr} &
      \multicolumn{4}{c|}{Youtube}\\
      & {Dim.} & {Micro$F_1$} & {Macro$F_1$}&{Comp.}  & {Dim.} & {Micro$F_1$} & {Macro$F_1$}&{Comp.}\\ 
      &            & {Score}            &{Score}            &{ time(s)}   &            & {Score}            &{Score}            &{ time(s)}\\
            \midrule
            \texttt{node2vec}& 128 & 37.58 & 21.05 & 44408.30 / 12.33 hr 
                             & 128 & 47.724 & 41.214 & 39133.25 /10.87 hr\\
            \texttt{deepwalk}& 128  & 41.2 & 29.61 & 7544.83 / 2.1 hr
                             & 128  & 46.82 & 39.37 & 100579.85 / 27.93 hr\\
            \texttt{LINE} &128  & 28.19 & 13.96 & 266814.56 / 74 hr
                              & 128 & 29.47 & 17.75 &  1966862 / 546 hr\\
            \texttt{VERSE} &128  & 41.52 & 31.41 & 17289.11 / 4.8 hr
                           & 128 & 47.80 & 41.44 & 253484.53 / 70 hr\\
            \texttt{NetMF} &128 & \texttt{OOM} & \texttt{OOM} & \NA 
                           & 128 &\texttt{OOM} & \texttt{OOM} & \NA\\
            \texttt{NodeSketch} &128 & 22.37 & 7.1  & 1057.30 / 17 min
                                &128 & 37.55 & 26.98 & 902.68 / 15 min\\
            \hline 
            \texttt{SGH}& 128 & \texttt{OOM} & \texttt{OOM} & \NA                                        & 128 & \texttt{OOM} & \texttt{OOM} &\NA\\
             \hline    
             \texttt{QUINT} & 100 & 19.59 & 5.4 & 20.86
                            & 100 & 28.66 & 16.65 & 17.90 \\
                            
                            & 128 & 21.08 & 7.3 & 23.11
                            & 128 & 30.38 & 20.10 & 20.46\\
                            
                            & 200 & 23.58 & 10.68 & 23.88
                            & 200 & 32.31 & 22.76 & 20.80\\
                            
                            & 256 & 25.41 & 12.46 & 25.80
                            & 256 & 32.67 & 24.10 & 26.74 \\
                            
                            & 1000 & 28.04 & 18.61 & 30.63              
                            & 1000 & 35.22 & 31.63 & 35.06\\
                            
                            & 2000 & 30.12 & 21.76 & 33.65              
                            & 2000 & 36.56 & 31.69 & 51.57\\
                            
                            & 3000 & 31.52 & 23.58 & 36.07              
                            & 3000 & 37.80 & 33.40 & 67.53\\
                            
                            & 4000 & 32.35 & 25.05 & 38.47              
                            & 4000 & 38.93 & 35.12 & 83.37\\
                            
                            & 6000 & 33.98 & 26.54 & 41.85           
                            & 6000 & 39.61 & 35.89 & 111.52  \\
                            
                            & 8000 & 34.66 & 27.31 & 44.86         
                            & 8000 & 40.94 & 37.06 & 140.08\\
                            
                            & 10000 & 35.08 & 28.10 &  46.87             
                            & 10000 & 41.20 & 37.46 & 167.94 \\
                            
                            & 16000 & 36.86 & 29.75 & 55.23          
                            & 16000 & 42.41 & 38.01 & 250.03\\
                            
                            & 20000 & 37.04 & 29.99 & 59.89              
                            & 20000 & 43.26 & 38.68 &  325.48\\
            \hline   
            \texttt{Uncompressed} &\NA  & \texttt{OOM} &   \texttt{OOM} & \NA     
            &\NA & \texttt{OOM} & \texttt{OOM} &  \NA\\

     \bottomrule
  \end{tabular}\\[1em]

  \begin{tabular}{|l|cccc| cccc |}
    \toprule
    \multirow{1}{*}{Method} &
      \multicolumn{4}{c|}{IMDB} &
      \multicolumn{4}{c|}{Pubmed}\\
      & {Dim.} & {Micro$F_1$} & {Macro$F_1$}&{Comp.}  & {Dim.} & {Micro$F_1$} & {Macro$F_1$}&{Comp.}\\ 
      &            & {Score}            &{Score}            &{ time(s)}   &            & {Score}            &{Score}            &{ time(s)}\\
            \midrule
            \texttt{node2vec} & 128 & 51.91 & 51.6  & 117 
                              & 128 & 43.2  & 31.34 & 5.88\\
            \texttt{deepwalk} & 128 & 50.68 & 50.34 & 118 
                              & 128 & 42.46 & 35.15 & 5.78\\
            \texttt{LINE} & 128 & 41.25 & 40.56 & 1850
                              & 128 & 58.19 & 53.81 & 840.25\\
            \texttt{VERSE} & 128 & 63.09 & 63.08 & 1194.10
                           & 128 & 42.29 & 31.93 & 806.98\\
            \texttt{NetMF} & 128 & 60.30 & 51.88 & 21.72
                           & 128 & 66.00 & 53.85 & 19.88\\
            \texttt{NodeSketch} & 128 & 55.69 & 46.64 & 6.98
                                & 128 & 43.79 & 33.77 & 4.31\\
            \hline 
            \texttt{SGH} & 128 & 70.83 & 70.75 & 110.78
                         & 128 & 51.10 & 44.80 & 67.67\\
                         
                          & 3000 & 74.51 & 74.49 & 174.12
                          & 2000 & 54.83 & 49.68 & 122.77\\
             \hline    
             \texttt{QUINT} & 100 & 40.37 & 39.78 & 0.72
                            & 100 & 47.71 & 39.27 & 0.2\\
                            
                            & 128 & 42.82 & 41.35 & 0.756 
                            & 128 &  48.63& 42.55 &0.21 \\
                        
                            & 200 & 46.91 & 46.48 & \textbf{0.76}
                            & 200 & 50.62 & 43.12 & \textbf{0.23}\\
                            
                            & 256 & 53.89 & 52.45 & 0.78 
                            & 256 & 52.96 & 47.58 & 0.25 \\
                           
                            & 1000 & 70.7 & 70.58 & 1.04
                            & 1000 & 63.28 & 59.63 & 1.23\\
                            
                            & 2000 & 83.88 & 83.85 & 3.89 
                            & 2000 & 68.27 & 65.59 & 2.50\\
                            
                            & 3000 & \textbf{89.76} & \textbf{89.81} & 6.32
                            & 3000 & 69.25 & 66.32 &3.75 \\
                            
            \hline   
            \texttt{Uncompressed} &\NA  & \textbf{94.90} &   \bf{95.91} &  \hspace*{0.25cm} \NA       \hspace*{0.25cm}
            &\NA &\textbf{76.52} & \textbf{74.98} &  \hspace*{0.25cm} \NA \hspace*{0.25cm}\\

     \bottomrule
  \end{tabular}\\[1em]
    \noindent\begin{tabular}{|l|cccc|cccc| }
    \toprule
    \multirow{1}{*}{Method} &
      \multicolumn{4}{c|}{Citeseer} &
      \multicolumn{4}{c|}{Cora} \\
      & {Dim.} & {Micro$F_1$} & {Macro$F_1$}&{Comp.}& {Dim.} & {Micro$F_1$} & {Macro$F_1$}&{Comp.}  \\ 
      &            & {Score}            &{Score}            &{ time(s)} &            & {Score}            &{Score}            &{ time(s)}\\
            \midrule
            \texttt{node2vec} & 128 & 29.77 & 21.90 & 1.75 
                              & 128 & 52.76 & 41.31 & 1.72\\
            \texttt{deepwalk}& 128  & 25.05 & 17.06 & 1.77
                             & 128  & 44.15 & 28.64 & 1.49\\
            \texttt{LINE} & 128 & 36.72 & 32.72 & 17.14
                              & 128 & 56.33 & 53.51 & 15.72\\
            \texttt{VERSE} & 128 & 32.93 & 29.18 & 71.83 
                           & 128 & 52.27 & 46.70 & 53.2\\
            \texttt{NetMF} & 128 & \textbf{49.69} & \textbf{44.06} & 1.39
                           & 128 & 57.68 & 52.65 & 1.43\\
            \texttt{NodeSketch} & 128 & 25.65 & 21.26 & 0.41
                                & 128 & 39.72 & 30.39 & 0.5\\
            \hline 
            \texttt{SGH} & 128 & 37.77 & 33.37 & 2.79
                         & 128 & 56.45 & 52.96 & 2.54\\
                        
                         & 2000 & 38.77 & 35.60 & 53.68
                         & 2000 & 57.93 & 56.07 & 36.33\\
             \hline    
             \texttt{QUINT} & 100 & 34.3 & 28.42 & 0.024
                            & 100 & 44.65 & 38.58 & 0.02\\
                            
                            & 128 & 34.95 & 28.75 & 0.029 
                            & 128 & 46.37 & 41.71 & 0.021\\
             
                            & 200 & 36.51 & 28.98 &  \textbf{0.035}
                            & 200 & 51.78 & 47.87 & \textbf{0.024} \\
                            
                            & 256 & 38.55 & 29.80 & 0.038
                            & 256 & 54.12 & 50.55 & 0.027 \\
                           
                            & 1000 & 44.16 & 38.79 & 0.076 
                            & 1000 & 61.62 & 60.43 & 0.055 \\
                           
                            & 2000 & 48.28 & 43.16  &  0.112 
                            & 2000 & \textbf{63.59} & \textbf{61.64} & 0.094\\
                            
                            & 3000 & 49.15 & 43.98 & 0.214  
                            & 3000 &  64.945 & 64.32 & 0.124 \\
                            
            \hline   
            \texttt{Uncompressed} &\NA &\textbf{51.40} &\textbf{47.06}  & \hspace*{0.25cm} \NA \hspace*{0.25cm}
                                  & \NA &\textbf{65.75} &\textbf{66.66} & \hspace*{0.25cm} \NA \hspace*{0.25cm}\\

     \bottomrule
  \end{tabular}\\ [1em]
  \vspace{-5mm}
\end{table*}
}\fi

\begin{table*}
  
\centering
\caption{Comparison on Micro $F_1$, Macro $F_1$ and compressions time of \texttt{QUINT} and other baselines (using 256 dimensions) for the task of node classification.
We stopped baselines that took 10 hours or more and indicate them by \DNS; {\tt OOM} indicates baselines that ran out of memory. Smaller datasets were embedded to a maximum of 3000 dimensions. 
The best Macro $F_1$ scores among the baselines and for {\tt QUINT} are {\bf bolded}.}
  \label{tab:node_classification_256}

   \noindent \begin{tabular}{|l|cccc| cccc |}
    \toprule
    \multirow{1}{*}{Method} &
      \multicolumn{4}{c|}{Flickr} &
      \multicolumn{4}{c|}{Youtube}\\
      & {Dim.} & {Micro$F_1$} & {Macro$F_1$}&{Comp.}  & {Dim.} & {Micro$F_1$} & {Macro$F_1$}&{Comp.}\\ 
      &            & {Score}            &{Score}            &{ time(s)}   &            & {Score}            &{Score}            &{ time(s)}\\
            \midrule
            \texttt{node2vec}& 256 & \NA[36.79] & \NA[20.28]  & \DNS[53138.81] 
                             & 256 & \NA[46.51]  & \NA[40.69] & \DNS[45643.165]\\
                             
            \texttt{deepwalk}& 256  & 41.26  & 30.06 & 16286
                             & 256  & \NA[46.55] & \NA[38.98] & \DNS[208670.25]\\
                             
            \texttt{LINE} &256  &  \NA[25.64] &  \NA[11.71] &  \DNS[605083.75]
                              & 256 & \NA[26.18]  &  \NA[15.35] & \DNS[4196271.8] \\
                              
            \texttt{VERSE} & 256  &  41.29 &  {\bf 30.95} &  23498.55
                           & 256 &  \NA[47.62] & \NA[41.06]  & \DNS[338381.29]\\

            \texttt{NetMF} &256 & \NA & \NA & \OOM 
                           & 256 & \NA & \NA & \OOM \\
                           
            \texttt{NodeSketch} &256 & 21.56 &  0.688 & 2530.36  
                                & 256 & 36.29  & {\bf 26.11} & 2233.81 \\
            \hline 
            \texttt{SGH} & 256 & \NA & \NA & \OOM           
                         & 256 & \NA & \NA & \OOM\\
             \hline    
             \texttt{QUINT} & 100 & 19.59 & 0.54 & 20.86
                            & 100 & 28.66 & 16.65 & 17.90 \\
                            
                            & 128 & 21.08 & 0.73 & 23.11
                            & 128 & 30.38 & 20.10 & 20.46\\
                            
                            & 200 & 23.58 & 10.68 & 23.88
                            & 200 & 32.31 & 22.76 & 20.80\\
                            
                            & 256 & 25.41 & 12.46 & 25.80
                            & 256 & 32.67 & 24.10 & 26.74 \\
                            
                            & 1000 & 28.04 & 18.61 & 30.63              
                            & 1000 & 35.22 & 31.63 & 35.06\\
                            
                            & 2000 & 30.12 & 21.76 & 33.65              
                            & 2000 & 36.56 & 31.69 & 51.57\\
                            
                            & 3000 & 31.52 & 23.58 & 36.07              
                            & 3000 & 37.80 & 33.40 & 67.53\\
                            
                            & 4000 & 32.35 & 25.05 & 38.47              
                            & 4000 & 38.93 & 35.12 & 83.37\\
                            
                            & 6000 & 33.98 & 26.54 & 41.85           
                            & 6000 & 39.61 & 35.89 & 111.52  \\
                            
                            & 8000 & 34.66 & 27.31 & 44.86         
                            & 8000 & 40.94 & 37.06 & 140.08\\
                            
                            & 10000 & 35.08 & 28.10 &  46.87             
                            & 10000 & 41.20 & 37.46 & 167.94 \\
                            
                            & 16000 & 36.86 & 29.75 & 55.23          
                            & 16000 & 42.41 & 38.01 & 250.03\\
                            
                            & 20000 & 37.04 & {\bf 29.99} & 59.89              
                            & 20000 & 43.26 & {\bf 38.68} &  325.48\\
            \hline   
            \texttt{Uncompressed} &\NA  & \texttt{OOM} &   \texttt{OOM} & \NA     
            &\NA & \texttt{OOM} & \texttt{OOM} &  \NA\\

     \bottomrule
  \end{tabular}\\[1em]

  \begin{tabular}{|l|cccc| cccc |}
    \toprule
    \multirow{1}{*}{Method} &
      \multicolumn{4}{c|}{IMDB} &
      \multicolumn{4}{c|}{Pubmed}\\
      & {Dim.} & {Micro$F_1$} & {Macro$F_1$}&{Comp.}  & {Dim.} & {Micro$F_1$} & {Macro$F_1$}&{Comp.}\\ 
      &            & {Score}            &{Score}            &{ time(s)}   &            & {Score}            &{Score}            &{ time(s)}\\
            \midrule
            \texttt{node2vec} & 256 & 49.16 & 50.81 & 138.79
                              & 256 & 41.39 & 29.99 & 6.77\\
                              
            \texttt{deepwalk} & 256 &47.33  & 46.12 & 256.61
                             & 256 & 42.51 & 35.23 & 12.29\\
                             
            \texttt{LINE} & 256 & 41.65 & 40.85 & 4205.38
                              & 256 & 55.21 &  50.64 & 1781.6 \\
                              
            \texttt{VERSE} & 256 & 61.56 & 60.74 & 1845.915
                           & 256 & 41.46 & 31.17  & 1070.63 \\
                           
            \texttt{NetMF} & 256 & 60.35 & 51.91 & 29.54
                          & 256 & 65.49 & {\bf 53.23} & 28.66 \\
                          
            \texttt{NodeSketch} & 256 & 54.86 & 44.27 & 17.05
                               & 256 & 43.27 & 33.14  & 9.924 \\
            \hline 
            \texttt{SGH} & 256 &  70.56 & {\bf 70.21} & 125.91
                         & 256 & 49.35  & 40.62  &74.54 \\
                         
             \hline    
             \texttt{QUINT} & 100 & 40.37 & 39.78 & 0.72
                            & 100 & 47.71 & 39.27 & 0.2\\
                            
                            & 128 & 42.82 & 41.35 & 0.756 
                            & 128 &  48.63& 42.55 &0.21 \\
                        
                            & 200 & 46.91 & 46.48 & 0.76
                            & 200 & 50.62 & 43.12 & 0.23\\
                            
                            & 256 & 53.89 & 52.45 & 0.78 
                            & 256 & 52.96 & 47.58 & 0.25 \\
                           
                            & 1000 & 70.7 & 70.58 & 1.04
                            & 1000 & 63.28 & 59.63 & 1.23\\
                            
                            & 2000 & 83.88 & 83.85 & 1.61
                            & 2000 & 68.27 & 65.59 & 2.50\\
                            
                            & 3000 & 89.76 & \textbf{89.81} & 2.1
                            & 3000 & 69.25 & {\bf 66.32} &3.75 \\
                            
            \hline   
            \texttt{Uncompressed} &\NA  & 94.90 &   95.91 &  \hspace*{0.25cm} \NA       \hspace*{0.25cm}
            &\NA & 76.52 & 74.98 &  \hspace*{0.25cm} \NA \hspace*{0.25cm}\\

     \bottomrule
  \end{tabular}\\[1em]
    \noindent\begin{tabular}{|l|cccc|cccc| }
    \toprule
    \multirow{1}{*}{Method} &
      \multicolumn{4}{c|}{Citeseer} &
      \multicolumn{4}{c|}{Cora} \\
      & {Dim.} & {Micro$F_1$} & {Macro$F_1$}&{Comp.}& {Dim.} & {Micro$F_1$} & {Macro$F_1$}&{Comp.}  \\ 
      &            & {Score}            &{Score}            &{ time(s)} &            & {Score}            &{Score}            &{ time(s)}\\
            \midrule
            \texttt{node2vec} & 256 &  28.13 &  21.53 & 2.03
                              & 256 & 51.22 &  40.56 & 1.98\\
                              
            \texttt{deepwalk}& 256 & 23.88 & 16.34 & 3.2
                             & 256 & 42.55 &  28.18 & 2.83\\
                             
            \texttt{LINE} &256 & 36.26 & 31.51 & 37.62
                              & 256 &56.25  & {\bf 53.37} & 34.85 \\

            \texttt{VERSE} & 256 & 33.04 & 29.43 & 109.62
                           & 256 & 50.15 & 45.34 &  64.17\\
                           
            \texttt{NetMF} & 256 & 49.75 &  {\bf 44.12} & 2.064
                           & 256 & 57.59 & 52.47 & 1.813\\
                           
            \texttt{NodeSketch} & 256 & 22.73 & 19.67 & 0.981
                                & 256 & 36.49 & 28.13 & 1.2\\
            \hline 
            \texttt{SGH} & 256 & 36.15 & 32.52 & 3.05
                         & 256 & 53.62 & 50.42 & 2.81\\

             \hline    
             \texttt{QUINT} & 100 & 34.3 & 28.42 & 0.024
                            & 100 & 44.65 & 38.58 & 0.02\\
                            
                            & 128 & 34.95 & 28.75 & 0.029 
                            & 128 & 46.37 & 41.71 & 0.021\\
             
                            & 200 & 36.51 & 28.98 &  0.035
                            & 200 & 51.78 & 47.87 & 0.024 \\
                            
                            & 256 & 38.55 & 29.80 & 0.038
                            & 256 & 54.12 & 50.55 & 0.027 \\
                           
                            & 1000 & 44.16 & 38.79 & 0.076 
                            & 1000 & 61.62 & 60.43 & 0.055 \\
                           
                            & 2000 & 48.28 & 43.16  &  0.112 
                            & 2000 & 63.59 & {61.64} & 0.094\\
                            
                            & 3000 & 49.15 & 43.98 & 0.214  
                            & 3000 &  64.945 & {\bf 64.32} & 0.124 \\
                            
                            & 3500 & 50.68 & {\bf 45.99} & 0.286
                            & & & & \\
            \hline   
            \texttt{Uncompressed} &\NA & 51.40 & 47.06  & \hspace*{0.25cm} \NA \hspace*{0.25cm}
                                  & \NA & 65.75 & 66.66 & \hspace*{0.25cm} \NA \hspace*{0.25cm}\\

     \bottomrule
  \end{tabular}\\ [1em]

  \vspace{-5mm}
\end{table*}

\begin{table*}
     \centering
 
 \caption{Performance evaluation of node classification on BlogCatalog}
 
 \begin{tabular}{|l|cccc|}
    \toprule
    \multirow{1}{*}{Method} &
      \multicolumn{4}{c|}{BlogCatalog} \\
      & {Dim.} & {Micro$F_1$} & {Macro$F_1$}&{Comp.} \\ 
      &            & {Score}            &{Score}            &{ time(s)}\\
            \midrule
            \texttt{node2vec} & 128 & 38.90 &  23.19 & 1359.25 \\
                              & 256 &	39.5171 &	23.2856 &	1607.86\\
            \hline           
            \texttt{deepwalk} & 128 & 41.82  & 28.24 & 1167.95\\
                             & 256 & 40.3932 &	27.8693	& 2457.24\\
             \hline               
            \texttt{LINE} & 128 & 20.17 & 10.15 & 1898.23\\
                              & 256 &	17.5174 &	9.3982 & 4349.56\\
             \hline                
            \texttt{VERSE} & 128 & 41.78 & 29.04  & 847.4 \\
                           & 256 & 40.55 & 28.05 & 1204.72\\
             \hline            
            \texttt{NetMF} & 128 & 43.45 & {\bf 29.05} &  72.39\\
                           & 256 & 43.98 & {\bf 30.905} & 101.26\\
              \hline            
            \texttt{NodeSketch} & 128 & 19.76 & 8.66 & 66.36 \\ 
                                & 256 & 19.0264 & 9.9705 & 175.527\\
                               
           \hline 
            \texttt{SGH} & 128 & 13.20   & 4.05 & 18.25  \\
                         & 256 & 10.6008 & 4.608 & 20.45\\
                         & 2000 & 8.09 & 4.42 & 111.50 \\
                         & 3000 & 9.02 & 4.74 & 120.27\\
             \hline    
             \texttt{QUINT} & 100 & 23.78 & 10.05 &  0.84\\
                            
                            & 128 & 23.74 & 11.87 & 0.88 \\
                        
                            & 200 & 25.86 & 13.73 & 0.96\\
                            
                            & 256 & 26.42 & 14.86 & 1.19\\

                            & 1000 & 31.66 & 20.32 & 1.55 \\

                            & 2000 & 34.69 & 21.50 & 1.59\\

                            & 3000 & 34.28 & 22.50 & 1.64\\
                            
                            & 4000 & 35.64 & 22.83 & 1.69\\
                            
                            & 5000 & 36.23 & 23.36 & 1.85\\
                            
                            & 6000 & 36.05 & 24.04 & 1.95\\
                            
                            & 8000 & 36.82 &24.95 & 2.13\\
                           
                            & 10000 & 37.70 & {\bf 25.51} & 2.35\\

            \hline   
            \texttt{Uncompressed} &\NA  & 38.2254 &   25.647 &  \hspace*{0.25cm} \NA       \hspace*{0.25cm}\\
    \midrule
 \end{tabular}
 \label{tab:blogcatalog_node_class}
 \end{table*}
 
\begin{table*}
\centering
 
\caption{Comparison on AUC-ROC score and compression time of \texttt{QUINT} and other baselines (using 128 dimensions) for the link prediction experiments.
We stopped baselines that took 10 hours or more and indicate them by \DNS; {\tt OOM} indicates baselines that ran out of memory. Smaller datasets were embedded to a maximum of 4000 dimensions. 
The best AUC-ROC among the baselines and for {\tt QUINT} are {\bf bolded}.
}

    \noindent \begin{tabular}{|l|ccc| ccc| ccc|}
    \toprule
    \multirow{1}{*}{Method} &
      \multicolumn{3}{c|}{Gowalla} &
      \multicolumn{3}{c|}{Flickr} &
      \multicolumn{3}{c|}{Youtube}\\
       & {Dim.} & {AUC-ROC} &{Time(s)}  & {Dim.} & {AUC-ROC} &{Time(s)}  & {Dim.} & {AUC-ROC} &{Time(s)}  \\

            \midrule
     \texttt{node2vec} & 128  & 75.48 & 7827  
                       & 128  & 76.4 & 8706.8 
                       & 128  & 65.1 & 30899.28 \\
        \texttt{deepwalk} & 128  & 74.79 & 7075
                          & 128  &  67.7    & 4978.64
                          & 128  &  \NA[55.4]    & \DNS[91783.87] \\

        \texttt{LINE}  & 128 & \NA[75.33] & \DNS[72242] 
                           & 128 &  \NA[41.5]   & \DNS[186796.55]
                           & 128 &  \NA[47.69]   & \DNS[1219860.63] \\

       \texttt{VERSE} & 128 & \textbf{87.35} & 4902.8 
                      & 128 & {\bf 91.6}  & 9469.08 
                      & 128 & \NA[87.04]  & \DNS[174982.02]  \\
                     
         
         \texttt{NetMF} & 128 & \NA  & \OOM
                        & 128 &  \NA & \OOM
                        & 128 &  \NA  & \OOM \\

         \texttt{NodeSketch} & 128  & 50.02 & 328.30 
                             & 128  & 50.03 & 770.65
                             & 128  & 50.04 & 538.81 \\
                            
         \hline
         
          \texttt{SGH} & 128 & \NA  & \OOM
                        & 128 &  \NA & \OOM
                        & 128 &  \NA  & \OOM \\

                       
         \hline
         \texttt{QUINT} & 100 & 84.93 & 4.2 
                        & 100 & {\bf 86.7} & 7.41
                        & 100 & {\bf 90.00} & 8.861 \\

                        & 128 & 84.63 & 4.38
                        & 128 & 86.67 & 7.69
                        & 128 & 89.64 & 9.05 \\

                        & 200 & 83.99 & 4.69  
                        & 200 & 86.2 & 8.47
                        & 200 & 88.6 & 9.31 \\

                        & 256 & 87.02 & 4.89 
                        & 256 & 86.14 & 9.18
                        & 256 & 87.94 & 9.45\\

                        & 1000 & {\bf 87.71} & 7.15 
                        & 1000 & 85.7 & 11.07
                        & 1000 & 83.3 & 17.77 \\
                        
                        & 2000 & 87.49 & 9.91 
                        & 2000 & 86.1 & 13.74
                        & 2000 & 80.7 & 24.61 \\

                        & 3000 & 87.28 & 12.58 
                        & 3000 & 86.4 & 14.83
                        & 3000 & 79.3 &  33.97\\

                        & 4000 & 87.15 & 15.28 
                        & 4000 & 86.6 & 15.5
                        & 4000 & 78.39 & 41.62 \\


                        
         \hline
         \texttt{Uncompressed}   & \NA & \texttt{OOM}   & \NA  
                                 & \NA & \texttt{OOM} & \NA  
                                 & \NA & \texttt{OOM} & \NA \\
                                 
    \bottomrule
  \end{tabular} \\[1em]

  \begin{tabular}{|l|ccc| ccc| ccc|}
    \toprule
    \multirow{1}{*}{Method} &
      \multicolumn{3}{c|}{Enron} &
      \multicolumn{3}{c|}{BlogCatalog} &
      \multicolumn{3}{c|}{Facebook} \\
         & {Dim.} & {AUC-ROC} &{Time(s)}  & {Dim.} & {AUC-ROC} &{Time(s)}  & {Dim.} & {AUC-ROC} &{Time(s)} \\

            \midrule
     \texttt{node2vec} 
                       & 128  & 67.13 & 560  
                       & 128  & 63.12 & 979.02
                       & 128  & 93.64 & 63.98\\
        \texttt{deepwalk} 
                          & 128  & 66.7  & 515.36
                          & 128  & 61.00    & 774.31
                          & 128  & 93.10 & 54.39\\
                          
        \texttt{LINE}   
                           & 128 & 75.87 & 4216
                           & 128 & 66.08    & 2379
                           & 128 & 83.41 & 250.25\\
                           
       \texttt{VERSE} 
                      & 128 & \textbf{97.20} & 1058.42 
                      & 128 & 74.24          & 351.38 
                      & 128 & 94.32          & 88.34\\
         
         \texttt{NetMF}   
                        & 128 & 83.98         & 62.29 
                        & 128 & 71.68         & 7.7 
                        & 128 & {\bf 94.86} & 2.02\\
                        
         \texttt{NodeSketch} 
                             & 128  & 49.99 & 44.07
                             & 128  & 50.00 & 75.53
                             & 128  & 50.00 & 10.69\\
         \hline
         
          \texttt{SGH} 
                       & 128 & 71.49        & 31.28
                       & 128 & 70.05        & 7.53
                       & 128 & 92.20        & 5.3\\
                       
         \texttt{}     
                       & 1600 & 68.00       &106.01
                       & 2400 & {\bf 75.36}       & 84.74
                       & 1800 & 88.89       & 50.32\\
         \hline
         \texttt{QUINT} 
                        & 100 & 84.23 & 0.55
                        & 100 & 81.60 & 0.88
                        & 100 & 81.69 & 0.075\\

                        & 128 & 84.65 & 0.58 
                        & 128 & 82.18 & 0.92
                        & 128 & 82.98 & 0.096\\

                        & 200 & 84.35 & 0.61
                        & 200 & 83.26 & 0.99
                        & 200 & 84.39 & 0.105\\

                        & 256 & 85.6 & 0.64
                        & 256 & 83.71 & 1.02
                        & 256 & 85.50 & 0.12 \\

                        & 1000 & 92.88 & 0.99
                        & 1000 & 84.38 & 1.28 
                        & 1000 & 93.85 & 0.18\\

                        & 2000 & 94.26 & 5.39
                        & 2000 & 85.36 & 1.45
                        & 2000 & 94.09 & 0.24 \\

                        & 3000 & 94.51 & 7.12 
                        & 3000 & 85.64 & 1.61
                        & 3000 & 94.26 & 0.32 \\

                        & 4000 & 94.75 & 9.05 
                        & 4000 & {\bf 86.29} & 1.77
                        & 4000 & 94.58 & 0.45\\
                        
                        & 35000 & {\bf 95.23} & 28.47 
                        & & &
                        & 4500 & {\bf 95.30} & 0.53 \\
                         

         \hline
         \texttt{Uncompressed}     
                                 & \NA & 95.42 & \NA  
                                 & \NA & 86.46 & \hspace*{0.32cm}\NA \hspace*{0.32cm} 
                                 & \NA & 95.44 & \hspace*{0.3cm} \NA\hspace*{0.3cm}\\
    \bottomrule
  \end{tabular} \\[1em]

  \label{tab:link_prediction_full_128}
\end{table*}

\begin{table*}
 
\centering
\caption{Comparison on AUC-ROC score and compression time of \texttt{QUINT} and other baselines (using 256 dimensions) for the link prediction experiments.
We stopped baselines that took 10 hours or more and indicate them by \DNS; {\tt OOM} indicates baselines that ran out of memory. Smaller datasets were embedded to a maximum of 4000 dimensions. 
The best AUC-ROC among the baselines and for {\tt QUINT} are {\bf bolded}.}

    \noindent \begin{tabular}{|l|ccc| ccc| ccc|}
    \toprule
    \multirow{1}{*}{Method} &
      \multicolumn{3}{c|}{Gowalla} &
      \multicolumn{3}{c|}{Flickr} &
      \multicolumn{3}{c|}{Youtube}\\
       & {Dim.} & {AUC-ROC} &{Time(s)}  & {Dim.} & {AUC-ROC} &{Time(s)}  & {Dim.} & {AUC-ROC} &{Time(s)}  \\

            \midrule
     \texttt{node2vec} & 256  & 74.22 &   9174.34
                       & 256  & 72.5 &  10210.63
                       & 256  & {\bf 65.4} &  35612\\
        \texttt{deepwalk} & 256  & 74.82 &   14235.79
                          & 256  & 66.25 & 9473.68 
                          & 256  & \NA[48.65] & \DNS[178874.8] \\

        \texttt{LINE}  & 256  & \NA[73.67] &  \DNS[159673.55] 
                           & 256  & \NA[39.78] &  \DNS[395958.42]
                           & 256  & \NA[43.58] &  \DNS[2725822.2]\\

       \texttt{VERSE}  & 256  & {\bf 86.88} &   7005.9
                       & 256  & {\bf 86.42} &  11416.7
                       & 256  & \NA[{\bf 82.27}] & \DNS[234021.3] \\
                     
         
         \texttt{NetMF} & 256 & \NA  & \OOM
                        & 256 &  \NA & \OOM
                        & 256 &  \NA  & \OOM \\

         \texttt{NodeSketch} & 256  & 50 &   770.3
                             & 256  & 49.99 &  1930.74
                             & 256  & 48.84 & 1265.8 \\
                            
         \hline
         
         \texttt{SGH} & 256 & \NA  & \OOM
                        & 256 &  \NA & \OOM
                        & 256 &  \NA  & \OOM \\

                       
         \hline
         \texttt{QUINT} & 100 & 84.93 & 4.2 
                        & 100 & {\bf 86.7} & 7.41
                        & 100 & {\bf 90.00} & 8.861 \\

                        & 128 & 84.63 & 4.38
                        & 128 & 86.67 & 7.69
                        & 128 & 89.64 & 9.05 \\

                        & 200 & 83.99 & 4.69  
                        & 200 & 86.2 & 8.47
                        & 200 & 88.6 & 9.31 \\

                        & 256 & 87.02 & 4.89 
                        & 256 & 86.14 & 9.18
                        & 256 & 87.94 & 9.45\\

                        & 1000 & {\bf 87.71} & 7.15 
                        & 1000 & 85.7 & 11.07
                        & 1000 & 83.3 & 17.77 \\
                        
                        & 2000 & 87.49 & 9.91 
                        & 2000 & 86.1 & 13.74
                        & 2000 & 80.7 & 24.61 \\

                        & 3000 & 87.28 & 12.58 
                        & 3000 & 86.4 & 14.83
                        & 3000 & 79.3 &  33.97\\

                        & 4000 & 87.15 & 15.28 
                        & 4000 & 86.6 & 15.5
                        & 4000 & 78.39 & 41.62 \\


                        
         \hline
         \texttt{Uncompressed}   & \NA & \texttt{OOM}   & \NA  
                                 & \NA & \texttt{OOM} & \NA  
                                 & \NA & \texttt{OOM} & \NA \\
                                 
    \bottomrule
  \end{tabular} \\[1em]

  \begin{tabular}{|l|ccc| ccc| ccc|}
    \toprule
    \multirow{1}{*}{Method} &
      \multicolumn{3}{c|}{Enron} &
      \multicolumn{3}{c|}{BlogCatalog} &
      \multicolumn{3}{c|}{Facebook} \\
         & {Dim.} & {AUC-ROC} &{Time(s)}  & {Dim.} & {AUC-ROC} &{Time(s)}  & {Dim.} & {AUC-ROC} &{Time(s)} \\

            \midrule
     \texttt{node2vec} 
                       & 256  & 66.34 &  653.92 
                       & 256  & 62.5 &  1158
                       & 256  & 91.59 &  74.09\\
        \texttt{deepwalk} 
                          & 256  & 64.59 &  1043.8 
                          & 256  & 58.42 & 1559.03 
                          & 256  & 92.78 &  108.1\\
                          
        \texttt{LINE}   
                           & 256  & 71.26 &  9132.6 
                           & 256  & 60.45 &  5053.17
                           & 256  & 81.13 & 567.77 \\
                           
       \texttt{VERSE} 
                       & 256  & {\bf 94.87} & 1629.15  
                       & 256  & 73.19 &  452.78
                       & 256  & 91.48 & 138.1 \\
         
         \texttt{NetMF}   
                        & 256  & 83.13 & 95.86  
                        & 256  & 71.32 &  10.29
                        & 256  & {\bf 93.65} & 2.563 \\
                        
         \texttt{NodeSketch} 
                             & 256  & 50 & 113.22  
                             & 256  & 49.99 & 192.11 
                             & 256  & 50.02 & 24.87 \\
         \hline
         
          \texttt{SGH} 
                       & 256  & 70.66 &  34.86 
                       & 256  & 65.56 &  7.92
                       & 256  &  90.49 &  5.658\\
         \texttt{}     
                       & 1600 & 68.00       &106.01
                       & 2400 & {\bf 75.36}       & 84.74
                       & 1800 & 88.89       & 50.32\\
         \hline
         \texttt{QUINT} 
                        & 100 & 84.23 & 0.55
                        & 100 & 81.60 & 0.88
                        & 100 & 81.69 & 0.075\\

                        & 128 & 84.65 & 0.58 
                        & 128 & 82.18 & 0.92
                        & 128 & 82.98 & 0.096\\

                        & 200 & 84.35 & 0.61
                        & 200 & 83.26 & 0.99
                        & 200 & 84.39 & 0.105\\

                        & 256 & 85.6 & 0.64
                        & 256 & 83.71 & 1.02
                        & 256 & 85.50 & 0.12 \\

                        & 1000 & 92.88 & 0.99
                        & 1000 & 84.38 & 1.28 
                        & 1000 & 93.85 & 0.18\\

                        & 2000 & 94.26 & 5.39
                        & 2000 & 85.36 & 1.45
                        & 2000 & 94.09 & 0.24 \\

                        & 3000 & 94.51 & 7.12 
                        & 3000 & 85.64 & 1.61
                        & 3000 & 94.26 & 0.32 \\

                        & 4000 & 94.75 & 9.05 
                        & 4000 & {\bf 86.29} & 1.77
                        & 4000 & 94.58 & 0.45\\
                        
                        & 35000 & {\bf 95.23} & 28.47 
                        & & &
                        & 4500 & {\bf 95.30} & 0.53 \\                         
                         

         \hline
         \texttt{Uncompressed}     
                                 & \NA & 95.42 & \NA  
                                 & \NA & 86.46 & \hspace*{0.32cm}\NA \hspace*{0.32cm} 
                                 & \NA & 95.44 & \hspace*{0.3cm} \NA\hspace*{0.3cm}\\
    \bottomrule
  \end{tabular} \\[1em]

  \label{tab:link_prediction_full_256}
\end{table*}

\end{document}